\documentstyle{mn}
\newif\ifAMStwofonts

\def\cm3{cm$^{-3}$}
\def\kms{km~s$^{-1}$}
\def\msunyr{M$_{\odot}$\,yr$^{-1}$}
\def\lsun{L/L$_{\odot}$}

\def\mdot{\dot{\rm M}}
\def\vinfty{v_{\infty}\,}

\ifoldfss

  \ifCUPmtlplainloaded \else
    \NewTextAlphabet{textbfit} {cmbxti10} {}
    \NewTextAlphabet{textbfss} {cmssbx10} {}
    \NewMathAlphabet{mathbfit} {cmbxti10} {} 
    \NewMathAlphabet{mathbfss} {cmssbx10} {} 
  \fi
  \ifAMStwofonts
    \ifCUPmtlplainloaded \else
      \NewSymbolFont{upmath} {eurm10}
      \NewSymbolFont{AMSa} {msam10}
      \NewMathSymbol{\upi}     {0}{upmath}{19}
      \NewMathSymbol{\umu}     {0}{upmath}{16}
      \NewMathSymbol{\upartial}{0}{upmath}{40}
      \NewMathSymbol{\leqslant}{3}{AMSa}{36}
      \NewMathSymbol{\geqslant}{3}{AMSa}{3E}

       \let\le=\leqslant
       \let\ge=\geqslant
    \fi
  \fi
\fi 

\ifnfssone
  \newmathalphabet{\mathit}
  \addtoversion{normal}{\mathit}{cmr}{m}{it}
  \addtoversion{bold}{\mathit}{cmr}{bx}{it}
  \newmathalphabet{\mathbfit} 
  \addtoversion{normal}{\mathbfit}{cmr}{bx}{it}
  \addtoversion{bold}{\mathbfit}{cmr}{bx}{it}
  \newmathalphabet{\mathbfss} 
  \addtoversion{normal}{\mathbfss}{cmss}{bx}{n}
  \addtoversion{bold}{\mathbfss}{cmss}{bx}{n}
  \ifAMStwofonts
    \ifCUPmtlplainloaded \else
      \UseAMStwoboldmath
      \makeatletter
      \new@mathgroup\upmath@group
      \define@mathgroup\mv@normal\upmath@group{eur}{m}{n}
      \define@mathgroup\mv@bold\upmath@group{eur}{b}{n}
      \edef\UPM{\hexnumber\upmath@group}
      \new@mathgroup\amsa@group
      \define@mathgroup\mv@normal\amsa@group{msa}{m}{n}
      \define@mathgroup\mv@bold\amsa@group{msa}{m}{n}
      \edef\AMSa{\hexnumber\amsa@group}
      \makeatother
      \mathchardef\upi="0\UPM19
      \mathchardef\umu="0\UPM16
      \mathchardef\upartial="0\UPM40
      \mathchardef\leqslant="3\AMSa36
      \mathchardef\geqslant="3\AMSa3E

       \let\le=\leqslant
       \let\ge=\geqslant
    \fi
  \fi
\fi 

\ifnfsstwo
  \DeclareMathAlphabet{\mathbfit}{OT1}{cmr}{bx}{it}
  \SetMathAlphabet\mathbfit{bold}{OT1}{cmr}{bx}{it}
  \DeclareMathAlphabet{\mathbfss}{OT1}{cmss}{bx}{n}
  \SetMathAlphabet\mathbfss{bold}{OT1}{cmss}{bx}{n}
  \ifAMStwofonts
    \ifCUPmtlplainloaded \else
      \DeclareSymbolFont{UPM}{U}{eur}{m}{n}
      \SetSymbolFont{UPM}{bold}{U}{eur}{b}{n}
      \DeclareSymbolFont{AMSa}{U}{msa}{m}{n}
      \DeclareMathSymbol{\upi}{0}{UPM}{"19}
      \DeclareMathSymbol{\umu}{0}{UPM}{"16}
      \DeclareMathSymbol{\upartial}{0}{UPM}{"40}
      \DeclareMathSymbol{\leqslant}{3}{AMSa}{"36}
      \DeclareMathSymbol{\geqslant}{3}{AMSa}{"3E}

       \let\le=\leqslant
       \let\ge=\geqslant
    \fi
  \fi
\fi 

\ifCUPmtlplainloaded \else
  \ifAMStwofonts \else 
    \def\upi{\pi}
    \def\umu{\mu}
    \def\upartial{\partial}
  \fi
\fi

\title[Neon abundances in WC stars]
{Quantitative analysis of WC stars: Constraints on neon abundances
from ISO/SWS spectroscopy}
\author[Luc Dessart et al.]
       {Luc Dessart$^1$\thanks{Present Address: D\'{e}partement de Physique, 
Universit\'{e} Laval and Observatoire du Mont M\'{e}gantic, Qu\'{e}bec, 
QC G1K 7P4, Canada; email: ldessart@orion.phy.ulaval.ca}, 
Paul A. Crowther$^1$, D. John Hillier$^2$, 
Allan J. Willis$^1$, \cr Patrick W. Morris$^{3,4}$\thanks{Present Address:
SIRTF Science Center/IPAC, California Institute of Technology, M/S 100-22, 1200
E. California Blvd, Pasadena, CA 91125, USA}, Karel A. van der Hucht$^4$
        \\
        $^1$: Dept. of Physics \& Astronomy, University College London, 
        Gower St., London, UK.\\
         $^{2}$: Dept. of Physics \& Astronomy, University of Pittsburgh, PA
         15260, USA.\\
         $^3$: Astronomical Institute ``Anton Pannekoek'', University of
               Amsterdam, NL-1098 SJ Amsterdam, the Netherlands\\
         $^4$: Space Research Organization Netherlands, Sorbonnelaan 2, 
               NL-3584 CA Utrecht, the Netherlands}
\date{Accepted/Received}

\pagerange{\pageref{firstpage}--\pageref{lastpage}}
\pubyear{1999}

\begin{document}

\maketitle

\label{firstpage}

\begin{abstract}
Neon abundances are derived in four Galactic WC 
stars -- $\gamma^{2}$ Vel (WR11, WC8+O7.5III), HD\,156385 (WR90, WC7), 
HD\,192103 (WR135, WC8), and WR146 (WC5+O8) -- using mid-infrared
fine structure lines obtained with ISO/SWS. Stellar parameters for 
each star are derived using a non-LTE model atmospheric code 
(Hillier \& Miller 1998) together with ultraviolet (IUE), 
optical (INT, AAT) and infrared (UKIRT, ISO) spectroscopy.
In the case of $\gamma^{2}$ Vel, we adopt results from 
De Marco et al. (2000), who followed an identical approach.

ISO/SWS datasets reveal the [Ne\,{\sc iii}] 15.5$\mu$m line in each
of our targets, while [Ne\,{\sc ii}] 12.8$\mu$m, [S\,{\sc iv}] 10.5$\mu$m
and [S\,{\sc iii}] 18.7$\mu$m are observed solely in $\gamma^2$ Vel.
Using a method updated from Barlow et al. (1988) to account for
clumped winds, we derive Ne/He=3--4$\times$10$^{-3}$ by number, 
plus S/He=6$\times$10$^{-5}$ for $\gamma^2$ Vel. Neon is highly
enriched, such that Ne/S in $\gamma^2$ Vel is eight times higher than 
cosmic values. However, observed Ne/He 
ratios are a factor of two times lower than predictions of current 
evolutionary models of massive stars.  An imprecise mass-loss and 
distance were responsible for the much greater discrepancy in 
neon content identified by Barlow et al.

Our sample of WC5--8 stars span a narrow range in $T_{\ast}$ (=55--71kK),
with no trend towards higher temperature at earlier spectral type, 
supporting earlier results for a larger sample by Koesterke \& Hamann (1995).
Stellar luminosities range from 100,000 to 500,000$L_{\odot}$, while
10$^{-5.1} \le \dot{M}/(M_{\odot} {\rm yr}^{-1}) \le 10^{-4.5}$, adopting
clumped winds, in which volume filling factors are 10\%. In all cases, wind
performance numbers are less than 10, significantly lower
than recent estimates. Carbon abundances span 0.08 $\le$ C/He $\le$ 0.25
by number, while oxygen abundances remain poorly constrained.



\end{abstract}

                        
\section{Introduction}

Understanding the physics of massive ($M_{\rm init}$$\ga$25\,$M_\odot$) 
stars, their atmospheres, radiation, and evolution is important for many 
aspects of astrophysics since their powerful winds affect the 
energy  and momentum balance of the interstellar medium (ISM). 
In particular, Wolf-Rayet (WR) stars provide crucial tests of 
nuclear reaction chains.

However, quantitative analysis of such stars, represents a 
formidable challenge. The assumptions of plane-parallel 
geometry and local thermodynamic equilibrium (LTE), which are often
adopted for lower luminosity stars, are inadequate.  
Nevertheless, the properties of a large sample of carbon sequence (WC-type) 
WR stars have now been quantitatively derived using detailed models, accounting
for non-LTE effects, spherical geometry and an expanding atmosphere
(Howarth \& Schmutz 1992; Koesterke \& Hamann 1995). However, recent studies 
have demonstrated that clumping (Moffat et al. 1988; Hillier 1991, 1996;
Schmutz 1997) and line blanketing (Schmutz 1997; Hillier \& Miller 1998) 
may have a significant effect on the derived physical properties of WR stars.

Overall, evolutionary predictions for massive stars (e.g. 
Meynet et al. 1994) are in good agreement with the observed properties
of Wolf-Rayet stars. van der Hucht \& Olnon (1985) derived a 
Ne/He ratio for $\gamma^{2}$\,Vel (WR11, HD\,68273, WC8+O7.5) from IRAS
space-based observations, which was found to be in good agreement with
expectations. However, Barlow et al. (1988, hereafter BRA88) identified 
a numerical flaw in these calculations and added new ground-based 
observations to reveal Ne/He=1.0$\pm$0.35$\times$10$^{-3}$, a factor 
of six lower than predicted. Is this discrepancy due to failure 
of evolutionary models, peculiarities of the $\gamma^{2}$\,Vel binary 
system, or 
incorrect assumptions for the stellar properties of the WC8 star?
The combination of Short Wavelength Spectrometer (SWS), Infrared Space 
Observatory (ISO) observations of a larger sample of WC stars, plus
recent progress in quantitative modelling of WR stars, should provide 
a definitive answer to this question.

Recent space-based spectroscopy of WR11, obtained with the
Infrared Space Observatory (ISO), was presented by van der Hucht et al. (1996).
who quoted excellent agreement of the fine-structure neon line fluxes 
with observations used by BRA88. Morris et al. (1998) subsequently
re-estimated Ne/He using contemporary information on the stellar distance 
(van der Hucht et al. 1997; Schaerer et al. 1997) and 
mass-loss rate (Stevens et al. 1996) of WR11, which revealed a considerable 
neon enrichment. Willis et al. (1997, hereafter Paper~I) also used 
ISO/SWS to observe WR146, another WC+O binary, again revealing significantly 
enriched neon. Recently, Morris et al. (2000) analysed ISO/SWS observations
of WR147 (WN8+OB), revealing neon, sulphur and calcium abundances in good
agreement with cosmic values, as expected by evolutionary models.

In this paper, we supplement results from Paper~I with other WC stars
for which ISO-SWS observations are available, HD\,156385 (WR\,90, WC7)
and HD\,192103 (WR135, WC8). We also
re-analyse WR146 (WC5+O8) using more sophisticated analysis techniques
and in the light of other observational evidence (Niemela et al. 1998;
Dougherty et al. 2000). A neon abundance is also re-derived for 
WR11, following recent spectroscopic results from De Marco et al. (2000).

The outline of the present work is as follows. New UV, optical and IR
observations of our programme stars are presented in Sect.~\ref{sect2},
with basic properties discussed in Sect.~\ref{sect3}. The spectroscopic
technique is introduced in Sect.~\ref{sect4}, and applied in Sect.~\ref{sect5}.
Our spectroscopic results are discussed in Sect.~\ref{sect6}, with neon and 
sulphur abundances derived in Sect.~\ref{sect7}. Finally, conclusions are reached
in Sect.~\ref{sect8}.

\begin{table}
\caption[]{Programme Galactic WC stars. Spectral types are obtained from
Crowther et al (1998), which follows Smith et al. (1990) except that
revised WCE criteria are adopted. Wind velocities are taken from 
(a) Prinja et al. (1990), (b) Willis et al. (1997), or (c) St Louis
et al. (1993)}
\label{table0}
\begin{tabular}{rrrrrl}
\hline
WR                 & HD & Other & $v_{\infty}$ &Ref.& Spectral \\
                   &    &       & km\,s$^{-1}$ &    & Type  \\
\hline             
11                 & 68273 & $\gamma^{2}$ Vel &1550&c & WC8+O7.5III \\ 
90                 & 156385&                  &2045&a & WC7      \\
%
%
135                & 192103& V1042 Cyg        &1405&a & WC8    \\
146                &       &HM19--3                  &2700&b & WC5+O8  \\
\hline
\end{tabular}
\end{table}

\begin{table}
\caption[]{Journal of optical and IR spectroscopic observations.
All SWS observations were obtained with the AOT06 scan.}
\label{table1}
\begin{tabular}{l@{\hspace{-2mm}}r@{\hspace{2mm}}l@{\hspace{2mm}}
l@{\hspace{2mm}}r}
\hline
WR & Epoch & Telescope      & Wavelength &   Exposure  \\
   &       & Inst.          & Range ($\mu$m)& Time (s) \\
\hline
11  & 17 May 1996 & ISO--SWS & 2.38--45.2  & 7,876 \\
90 &  9 Mar 1998 & AAT--RGO    & 0.50-1.03   & 5     \\
   & 16 Feb 1997 & ISO--SWS & 2.60--19.6  & 10,068 \\
135&    Sep 1991 & INT--IDS   & 0.38--0.73  & 24    \\
   & 20 Aug 1994 & UKIRT--CGS4 & 1.03--1.13  &  64    \\
   & 19-21 Aug 1994 & UKIRT--CGS4 & 1.61--2.21  & 576 \\ 
   & 21 Aug 1994 & UKIRT--CGS4 & 2.30--2.51  & 128       \\ 
   & 11 Nov 1996 & ISO--SWS & 2.38--45.2  &  6,538 \\ 
146& 21 Jul 1996 & INT--IDS   & 0.36--0.68  & 800       \\
   & 19 Aug 1994 & UKIRT--CGS4 & 1.03--1.13  &  64    \\
   & 20 Aug 1994 & UKIRT--CGS4 & 1.23--1.33  &  64    \\
   & 19-21 Aug 1994 & UKIRT--CGS4 & 1.61--2.21  & 224 \\ 
   & 21 Aug 1994 & UKIRT--CGS4 & 2.30--2.51  & 64       \\
   & 12 May 1996 & ISO--SWS & 2.60--19.6  & 16,922 \\
\hline
\end{tabular}
\end{table}

\section{Observations}\label{sect2}

The programme Galactic WC stars are listed in Table~\ref{table0}
where we give the various catalogue names and our adopted
spectral types, following Smith et al. (1990) and Crowther et al. (1998).
We will refer to our programme stars by their WR catalogue number (van 
der Hucht et al. 1981). Wind velocities are taken from UV resonance
line measurements (Prinja et al. 1990; St Louis et al. 1993), 
except for WR146 for which a terminal velocity was obtained in 
Paper~I from ISO/SWS spectroscopy. In this section, we discuss the 
ground-based  (AAT, INT, ESO) and space-based  (ISO, IUE) observations, the 
journal  of which is presented in Table~\ref{table1}.

\subsection{Ultraviolet spectroscopy}

Our UV dataset was obtained solely from the International Ultraviolet 
Explorer (IUE) archive. Available high dispersion (HIRES), large 
aperture, short-(SWP) and long-wavelength (LWP, LWR) observations of 
WR90 and WR135 were combined to provide high S/N datasets (St Louis
1990). Absolute flux calibration was achieved following the calibration 
curve obtained by
Howarth \& Phillips (1986). Agreement between our final HIRES datasets
and archival low resolution (LORES) observations was found to be excellent.

Willis et al. (1986) have discussed the UV spectral morphology of WC
stars, including WR90 and WR135. The principal spectral features for
both stars are Si\,{\sc iv} $\lambda\lambda$1393-1407, 
C\,{\sc iv} $\lambda\lambda$1548--51, He\,{\sc ii} $\lambda$1640, and 
C\,{\sc iii} $\lambda$2297. A forest of Fe\,{\sc v-vi} lines are 
present in both stars, with Fe\,{\sc iv} also prominent in WR135.

\subsection{Optical spectroscopy}

Previously unpublished spectroscopic observations of WR90
and WR135 were obtained at the 2.5m Isaac Newton Telescope (INT) and 3.9m 
Anglo Australian Telescope (AAT).
The INT dataset was obtained with the IDS, 500mm camera, and
GEC CCD during 1991 September, while the AAT dataset 
was obtained with the RGO spectrograph, 25cm camera, and
a MIT/LL CCD during 1998 March. Each dataset was obtained using 
CCDs and reduced in a standard manner using software available on 
STARLINK. Spectra were debiased, flat-fielded, optimally extracted,
wavelength calibrated, and flux calibrated in the case of the INT dataset. 
Flux calibration was achieved for the AAT observations via scaling
to the level of the blue CTIO spectrophotometry taken 
from Torres-Dodgen \& Massey (1988).

Arc spectra provided a measure of  the instrumental resolution: 
2--3\AA\ (INT) and  4\AA\ (AAT). 
Complete spectral coverage between 3800--7300\AA\ (INT) and 5000--10300\AA\ 
(AAT) was achieved with, respectively, six and one grating settings.
The atmospheric absorption bands removed using suitable comparison 
stars. 


The optical spectral appearance of WR146 has previously been discussed in
Paper~I, while Dougherty et al. (2000) present a new high quality blue optical
observation of that star. The morphologies of WR90 and WR135 are relatively
similar, dominated by He\,{\sc i-ii} and C\,{\sc iii-iv} emission features,
except that the emission lines of WR90 are substantially broader. 
C\,{\sc iii} $\lambda\lambda$4747--51 and C\,{\sc iv}
$\lambda\lambda$5801--12 are the two strongest features in each case.
Several O\,{\sc iii-v} features are present around 
$\lambda\lambda$2950--3150, and $\lambda$5592 with C\,{\sc ii} present 
at $\lambda$4267 for WR135.

\subsection{Near-IR spectroscopy}

Our principal near-IR dataset was obtained at the 3.8m U.K. Infrared Telescope (UKIRT) with the 
cooled grating spectrograph CGS4, the 300mm camera, a 75\,l/mm grating and a
62$\times$58 InSb array in 1994 August. Observations of WR135 and WR146
were bias-corrected, flat-fielded, extracted and sky-subtracted
using {\sc cgs4dr} (Daly \& Beard 1992). Subsequent reductions and analysis
were carried out using {\sc figaro} (Shortridge et al. 1999) and {\sc dipso}
(Howarth et al. 1998). In order to remove atmospheric features, 
the observations were divided by
an appropriate standard star (whose spectral features were artificially
removed) observed at around the same time and similar air mass. 
In regions of low atmospheric transmission at UKIRT 
the reliability of line shape and strength 
must be treated with caution (e.g. the P$\alpha$ region).
The near-IR morphology of WR135 and WR146 have been discussed
previously by Eenens, Williams \& Wade (1991).

We also utilise intermediate dispersion CCD spectra of WR135
covering 0.97--1.03$\mu$m, published by Howarth \& Schmutz
(1992), and obtained at the INT with 
the Intermediate Dispersion Spectrograph (IDS) and a GEC CCD in 
1990 October.  Although Howarth \& Schmutz (1992) obtained observations 
of the He\,{\sc i} $\lambda$10830+P$\gamma$ line, we prefer to 
use our new, lower resolution UKIRT observations of this feature.
(The sharp fall in efficiency of the GEC CCD longward of 
P$\gamma$ leads to an ill-defined continuum.). We attempt an approximate
flux calibration for these data, by setting the local continuum 
via interpolation of the observed 0.7$\mu$m (INT) and 1.03$\mu$m (UKIRT)
flux levels.


\subsection{Mid-IR spectroscopy}

New mid-IR data of WR90 and WR146 were obtained as part 
of Guest Observer programme, AJWWOLF (P.I. Willis), with the 
Short Wavelength Spectrograph (SWS; de Graauw et al. 1996) onboard 
the ESA Infrared Space Observatory (Kessler et al. 1996).
Observations of HD\,117297 (WR53), HD\,164270 (WR103) and HD\,165763 (WR111) from 
the AJWWOLF programme were of insufficient quality to 
provide mid-IR line flux measurements, and so these  were excluded 
from the present study. Consequently, we have included 
SWS observations of WR11 and WR135 from the Guaranteed Time 
programme WRSTARS (P.I. van der Hucht). In all cases the 
SWS AOT6 observing mode was
used to achieve full grating resolution, $\lambda/\Delta\lambda \;
\sim \; 1300-2500$. The continuous wavelength coverage was 
2.60--19.6\,$\mu$m for datasets from the AJWWOLF programme, and 
2.38--45.0$\mu$m for WR135 and WR11.

Data of detector ``bands'' 1 and 2 
respectively cover wavelengths of 2.38--4.08\,$\mu$m using 12 In:Sb 
detectors and 4.00--12.05\,$\mu$m with 12 Si:Ga detectors, employing entrance 
slits that give an effective aperture area of 14~$\times$~20 arcsec on 
the sky.  Band 3A to 3D data cover 12.0--27.6\,$\mu$m using 12 
Si:As detectors, with sky coverage of 14~$\times$~27  arcsec. finally, band 
3E and 4 data cover 27.5--45.0$\mu$m using 12 Ge:Be detectors, with sky 
coverage of 20$\times$33 arcsec.

The total integration time was set to allow one complete scan over the 
wavelengths selected within each ``AOT band'', defined by the permissible 
combinations of detector band, aperture, and spectral order (cf. de Graauw 
et al. 1996).  The observation time includes dark current measurements, and 
a monitor of photometric drift for the detectors of bands 2--3.  A 
drift measurement is not normally made for the relatively stable In:Sb 
detectors.

The data processing of the SWS data for WR146 was discussed in Paper~I, with
WR11, WR90 and WR135 reduced in a similar manner. The stellar spectrum
of WR11 will be discussed elsewhere (Morris et al. 2000, in prep), with solely
mid-IR fine structure line fluxes measured here.



\begin{table}
\caption[]{Summary of photometry, reddenings and distances for programme 
stars, including comparisons with reddenings from the literature (lit). 
Observed magnitudes correspond to the systemic value for binary 
systems, and include the Schmutz \& Vacca (1991) corrections to 
Smith (1968) photometry ($\ast$)}
\label{table2}
\begin{tabular}{r@{\hspace{2.5mm}}l@{\hspace{2.5mm}}c
@{\hspace{2.5mm}}c@{\hspace{2.5mm}}c@{\hspace{2.5mm}}c@{\hspace{2.5mm}}c
@{\hspace{2.5mm}}c@{\hspace{2.5mm}}c}
\hline
WR               &$v^{\rm sys}$      &Ref. &$E_{\rm B-V}$ (lit)& R &
Ref.&$M_{v}^{\rm WC}$&$d$&Ref\\
                   & mag    &     & mag          & &  &mag&
 kpc& \\
\hline             
11             &  1.70$^{\ast}$  &a &0.04 (0.03)&3.1&e&$-$3.7& 0.26 & h    \\ 
90             &  7.41$^{\ast}$  &a &0.38 (0.44)&3.1&d&$-$4.7& 1.55 & e     \\
135            &  8.51  &b &0.37 (0.35)&3.1&a&$-$4.3& 2.09 & f \\ 
146            & 13.91  &c &2.87 (2.80)&2.9&c&$-$5.3& 1.40 & g \\ 
\hline
\end{tabular}
\begin{flushleft}
a: Smith (1968), b: Massey (1984), c: Paper~I, d: Morris et al. (1993)
e: Schaerer et al. (1998), f: Lundstr\"{o}m \& Stenholm (1984), 
g: Dougherty et al. (2000), h: van der Hucht et al. (1997)
\end{flushleft}
\end{table}
%
%

\section{Interstellar reddening and distances}\label{sect3}

In this section, interstellar reddenings and distances to the programme
stars are discussed. Pre-empting results from Sect.~\ref{sect4}, 
reddenings are directly obtained from comparing theoretical synthetic 
spectra with de-reddened observations. Distances either follow from
cluster/association membership, or assumed absolute magnitudes.
Table~\ref{table2} provides a summary of the derived reddenings and 
distances for our sample of stars. A comparison with reddenings 
from the recent literature shows good agreement.
The distances to WR135 (Cyg~OB3 member) and WR11 ({\sc hipparcos}) 
are known with confidence, while those adopted here for WR90 
and WR146 deserve comment.

\subsection{WR\,90}

WR90 is not a member of an association or cluster. We therefore
estimated its distance based on the mean absolute visual magnitude 
of other Galactic WC7 stars. Unfortunately, all five WC7 stars that are 
members of associations/clusters are within binary systems (Lundstrom
\& Stenholm 1984). We restricted the sample to those for which reliable
reddenings were known (HD97152=WR42, HD152270=WR79 and HD192641=WR\,137)
from Morris et al. (1993). An  absolute visual magnitude of 
$M_{\rm v}=-$4.7 mag($\sigma$=0.8) was obtained for the WC7 components
by comparing their emission line strengths with WR90 (specifically 
C\,{\sc iii-iv} $\lambda$\,4650, C\,{\sc iv}$\lambda$\,5804).
Our estimate is in good agreement with van der Hucht et al. (1988)
and Smith et al. (1990) who
derived $M_{\rm v}=-$4.8 mag. Pre-empting results from Sect.~\ref{sect5}
we derive $E_{B-V}$=0.38 mag, so that the uncertainty in the 
calibration and reddening ($\pm$\,0.02 mag) implies 
$d$=1.55\,kpc$^{+0.5}_{-0.4}$. 

\subsection{WR\,146}\label{wr146_dist}

Dougherty et al. (1996) used IR and mm photometry to estimate a 
distance  of 1.2$\pm$0.3\,kpc towards WR146. Radio observations
revealed two separate components, namely the (thermal) WC emission 
plus the (non-thermal) bow shock emission between the two winds. 

Meanwhile, a lower distance of 0.75$\pm$0.15\,kpc was
derived in Paper~I, using the mean $M_{v}$ for a WCE star (Smith 
et al. 1990), plus an assumed spectral type of O8.5\,V for the companion.
Subsequently, Niemela et al. (1998) used WFPC2 aboard the Hubble 
Space Telescope (HST) to measure the individual UBV magnitudes of 
the WC and OB components of WR146. They derived V=13.64 mag
for the WC star, 0.24 mag brighter than the OB component.
In contrast, a difference of 0.8 mag was assumed in Paper~I. 

The combined HST and radio datasets indicate that the OB companion 
possesses a powerful stellar wind, with a giant or supergiant 
luminosity class. Recent optical spectroscopy supports a supergiant 
O8 classification (Dougherty et al. 2000). As discussed by Dougherty
et al., a large distance to this system would result if the
absolute magnitude of the companion was typical of late O supergiants
(Conti \& Alschuler 1971). Instead, we adopt the distance of 
1.4kpc, as derived by Dougherty et al. At this distance, WR146 would 
be a foreground object to Cyg~OB2 (Lundstrom \& Stenholm 1984; 
Torres-Dodgen et al. 1991).

We derived $E_{\rm B-V}$=2.87 mag and R=2.9
using the optical-IR reddening law of Steenman \& Th\'{e} (1989, 1991). This
provided a superior comparison between theoretical predictions and de-reddened
observations than with either the Howarth (1983) or Cardelli et al. (1989) laws.
Therefore, $M_{v}=-$5.3 mag for the WCE component, by far in excess of 
`typical' WCE stars (Smith et al. 1990).

\begin{table}
\begin{center}
\caption[]{Summary of the WC model atoms, and source of atomic datasets.
N$_{\sc f}$ is the number of full levels, N$_{\sc s}$ the number 
of super levels and N$_{\sc trans}$ the corresponding number of 
transitions. The last column refers to the upper level of a given ion 
included in our treatment}
\label{tabatom}
\begin{tabular}{l@{\hspace{0.5mm}}r@{\hspace{2mm}}r@{\hspace{2mm}}r@{\hspace{4mm}}l@{\hspace{4mm}}l}
\hline
 Species          &      $N_{\sc f}$  &  $N_{\sc s}$ & $N_{\sc trans}$ & Ref & Details\\
\hline 
 He {\sc i}        &       39  &  27  &315 &a& $n\le$14\\
 He {\sc ii}       &       30  &  13  &435 &b& $n\le$30 \\
 He {\sc iii}      &       1   &    1 &    & &          \\  
 C {\sc ii}       &        88  &  39  &791 &c,d&$nl\le$2p3d $^{4}$D$^{o}$\\ 
 C {\sc iii}       &      243  &  99  &5513&e,f&$nl\le$2p4d $^{1}$D$^{o}$\\ 
 C {\sc iv}        &       64  &  49  &1446&g&$n\le$30\\
 C {\sc v}         &        1  &   1  &    & &\\ 
 O {\sc ii}        &        3  &   3  & 3  &e,h&$nl\le$2p$^{3}$ $^{2}$P$^{o}$ \\
 O {\sc iii}       &       50  &  50  &213 &e,i&$nl\le$2p4f $^{1}$D\\ 
 O {\sc iv}        &       72  &  30  &835 &e,j&$nl\le$2p3p'' $^{2}$P$^{o}$ \\
 O {\sc v}         &      91  &  31   &748 &e,k&$nl\le$2p4p $^{3}$P\\
 O {\sc vi}        &       19  &  13  & 72 &g&$n\le$5\\
 O {\sc vii}       &        1   &  1  &    & &\\ 
 Si {\sc iv}        &       28  &  17 &129 &h&$n\le$6\\
 Si {\sc v}         &        1  &   1 &    & &\\ 
 Fe {\sc iv}       &      280  &  21  &5055&l&$nl \le$3d$^4$($^{1}$G)4p$^{2}$P$^{\circ}$\\
 Fe {\sc v}        &      182  &  19  &2517&m&$nl \le$3d$^3$($^{2}$D)4p$^{1}$P$^{\circ}$\\
 Fe {\sc vi}       &       80  &  10  &722 &n&$nl \le$3d$^2$($^{1}$S)4p$^{2}$P$^{\circ}$\\
 Fe {\sc vii}      &      153  &  14  &1213&o&$nl \le$3p$^5$($^{2}$P)3d3(b$^{2}$D$^{1}$P$^{\circ}$ \\
 Fe {\sc viii}     &       1   &  1   &    & &   \\
\hline
                   &1299     & 440 & 20007     \\
\hline
\end{tabular}
\end{center}
(a) Fernley  et al. (1987); (b) Wiese et al. (1966); (c) Yu Yan et al. (1987);
(d) Yu Yan \& Seaton (1987); (e) Nussbaumer \& Storey (1983, 1984); 
(f) P.J.~Storey (unpublished); (g) Peach et al. (1988); (h) Seaton (1995); 
(i) Luo et al. (1989); (j) Luo \& Pradhan (1989); (k) Tully et al. (1990)
(l) Becker \& Butler (1995b); 
(m) Becker \& Butler (1992); (n) Becker \& Butler (1995a); (o) K.~Butler
(unpublished)
\end{table}

\section{Modelling technique}\label{sect4}

Before discussing the fine analysis of each programme star, we
introduce the spectroscopic technique followed here. We use the non-LTE 
code of Hillier \& Miller (1998) which iteratively solves the transfer equation in the co-moving frame 
subject to statistical and radiative equilibria in an expanding, spherically symmetric and steady-state atmosphere. Relative to earlier versions of this code (Hillier 1987, 1990), two major enhancements 
have been incorporated, of particular relevance to WC-type stars, namely (i)
line blanketing, (ii) clumping. Specific details of the 
techniques used are provided by Hillier \& Miller (1998, 1999), with only 
a brief overview given here. 

As discussed by Hillier \& Miller (1998, 1999), extremely complex atomic models
are necessary for the quantitative analysis of WC stars, a
computationally demanding requirement. Consequently, we make use of the 
technique of `super-levels', in which several atomic levels of similar 
energies and
properties are combined into a single one, a super level, with the populations
of the super level calculated in the rate equations. Populations of individual
atomic levels are then calculated by assuming that it has the same departure
coefficient as the corresponding super level to which it belongs.  In this way,
extremely complex atoms of helium, carbon, oxygen and iron can be considered.
In Table~\ref{tabatom} a total of 1299 full ($N_{\rm F}$) 
atomic levels are combined to produce just 440 super levels ($N_{\rm s}$). 
The number of iron transitions is limited by $gf \le 10^{-4}$, so that a total
of 20,007 transitions are included. In addition to the specific
models discussed here, we have also considered cases with expanded atomic 
datasets and ionization stages (e.g. Fe\,{\sc iii}), for 
which the differences in emergent synthetic spectrum were negligible.

Oscillator strengths, collision and photoionization cross-sections are 
taken from a wide variety of sources (Table~\ref{tabatom}). 
The OPACITY project (Seaton 1987, 1995) 
formed the basis of most radiative rates, supplemented by calculations
for CNO by Nussbaumer \& Storey (1983, 1984) and Storey (unpublished), 
and for iron by Becker \& Butler (1992, 1995ab) and Butler (unpublished).

The stellar radius ($R_{\ast}$) is defined
as the inner boundary of the model atmosphere and is located at
Rosseland optical depth of 10 with the stellar temperature ($T_{\ast}$)
defined by the usual Stefan-Boltzmann relation. Similarly, 
$T_{\rm 2/3}$ relates to the radius ($R_{\rm 2/3}$)
at which the Rosseland optical depth equals 2/3.

Although the majority of spectral analyses of WR stars have adopted a standard
$\beta$=1 velocity law, there is both observational and theoretical evidence
for a more slowly accelerating outflow (e.g. Schmutz 1997; Lepine \& 
Moffat 1999). Consequently, we
adopt a form for the velocity law (Eqn 8 from Hillier \& Miller 1999) 
such that two exponents are considered with the result that acceleration 
is modest at small radii, but continues to large distances, i.e.

 \[ v(r)=\frac{v_{0}+(\vinfty-v_{\rm ext}-v_{0})(1-R_{\star}/r)^{\beta_{1}}+
                                   v_{\rm ext}(1-R_{\star}/r)^{\beta_{2}}}
             {1+(v_{0}/v_{\rm core})(\exp((R_{\star}-r)/h_{\rm eff}))} \]

Here $v_{ext}$ is an intermediate terminal velocity, 
$v_{\rm core}$ the core velocity (typically a few km\,s$^{-1}$), $v_{0}$ is the 
photospheric velocity (typically 100.0 km\,s$^{-1}$), $\vinfty$ is the final
wind velocity and $h_{\rm eff}$ the scale height ($\sim$0.01$R_{\star}$). For all 
models, we take $\beta_{1}$=1 and $\beta_{2}$=50 as in Hillier \& Miller (1999).

There is now overwhelming evidence for the clumped nature of WR stars
(e.g., Moffat et al. 1988; Moffat 1999), 
so we have adopted a simple filling factor approach. 
We assume that the wind is clumped with a volume filling factor, $f$, and that
there is no inter clump material. Since radiation instabilities are not expected
to be important in the inner wind we parameterise the filling factor 
so that it approaches unity at small velocities. 
Clumped and non-clumped spectra are very similar, except that line profiles 
are slightly narrower with weaker electron 
scattering wings in the former. Although 
non-clumped models can be easily rejected, because of the severe line 
blending in WC winds $\dot{M}/\sqrt{f}$ is derived by our spectroscopic 
analysis, rather than $\dot{M}$ and $f$. Also, this formulation introduces a 
revision of the BRA88 analytical formula, which will be addressed in 
Sect~\ref{sect5}.

\section{Spectroscopic analysis of individual stars}\label{sect5}

In this section, we discuss the analysis of our programme WC stars. 
Stellar parameters ($T_{\ast}$, log\,$L/L_{\odot}$, $\dot{M}/\sqrt{f}$, C/He, O/He) were adjusted until the observed ionization balance, line strengths
and de-reddened optical continuum flux distribution were reproduced. 
Because of the substantial
effect  that differing mass-loss rates, temperatures and elemental abundances have 
on the emergent spectrum, this was an iterative process, in which initial
model parameters were adopted from the literature if available (e.g. 
Koesterke \& Hamann 1995; Morris et al. 1993). Terminal wind velocities
are tabulated in Table~\ref{table0}.

\begin{figure*} 
\vspace{23cm}
\includegraphics{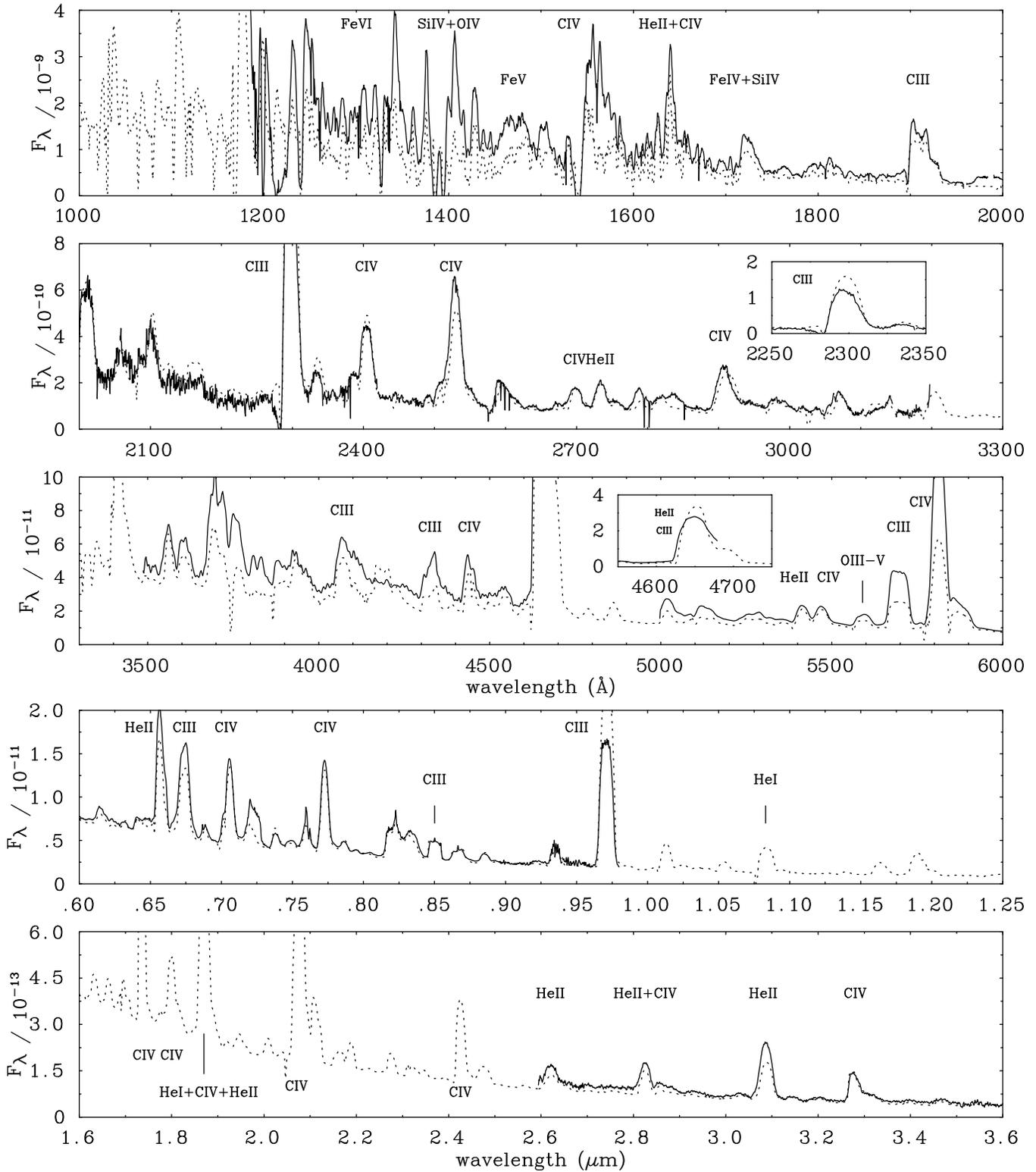}
\caption{Comparison between de-reddened ($E_{\rm B-V}$=0.38, R=3.1)
spectrophotometry of HD\,156385 = WR90 (WC7) obtained from
IUE, CTIO, AAT and ISO (solid lines), and line-blanketed, clumped model 
predictions (dotted lines). Stellar parameters are 
$\mdot$=2.5$\times$10$^{-5}$\msunyr,
$T_{\star}$=$71~kK$, log\,$L/L_{\odot}=5.5$, C/He=0.25 and O/He=0.03 by
number. Flux units are in ergs/cm$^{2}$/s/\AA }
\label{fig1}
\end{figure*}

The wind ionization balance was ideally selected on the basis of isolated
optical lines from adjacent ionization stages of carbon and/or helium.
In practice this was difficult to achieve because of the severe blending in
WC winds. The low wind velocity of WR135 allowed  He\,{\sc i} $\lambda$5876 
and He\,{\sc ii} $\lambda$5412 to be used as  ionization balance constraints. 
In other cases, the
wind ionization balance was selected on the basis of carbon diagnostics. 
Unfortunately, from all the possible WC diagnostics, the usual classification lines C\,{\sc iii}
$\lambda$5696/C\,{\sc iv} $\lambda$5804 are difficult to match,
being so sensitive to minor stellar parameter changes. 
Consequently alternative 
diagnostics were sought. For WR90, 
our primary diagnostics were C\,{\sc iii} $\lambda$8500/C\,{\sc iv} 
$\lambda$7736 since these suffered from negligible contamination and are
predicted by the model to vary smoothly across the temperature space.
In the case of WR146 this
spectral region, as well as lines of low ionization, were not available.
We selected He\,{\sc ii} $\lambda$5412/He\,{\sc i} 
$\lambda$10830, and C\,{\sc iii} $\lambda$6740/C\,{\sc iv} 1.74$\mu$m
as our primary ionization diagnostics.

Our experience with a variety of stellar models led us to select the strong
ultraviolet P Cygni profiles of carbon (C\,{\sc iii} $\lambda$1909, 2297)
 as our principal mass-loss diagnostics. A major limitation with 
C\,{\sc iv} $\lambda$1550 was that different mass-loss rates also affected 
nearby Fe lines, which strongly modulated the predicted strength of the 
emergent P Cygni profile. Since ultraviolet observations of WR146 are 
unavailable, we relied on He\,{\sc i} $\lambda$10830 as the principal 
mass-loss diagnostic in that case. 

 As in other recent spectroscopic studies of WC stars, He\,{\sc ii} $\lambda$5412/C\,{\sc iv} $\lambda$5471 were selected as the diagnostics for
C/He determinations since the relative strength of  these features are insensitive to differences of temperature or mass-loss rate.  As discussed
by Hillier \& Miller (1998), this spectral region also contains a number of
additional weak features, common to all stars. More problematic is that
misleading C/He ratios would be obtained if (i) a limited number of C\,{\sc iv}
atomic levels were included; (ii) homogeneous models were adopted in which
electron scattering wings were incorrectly predicted.

Test calculations were also performed using recombination theory.
Comparisons of C/He with our more sophisticated results for WCE stars 
was found to be reasonable -- we obtain C/He=0.08 by number for WR146 
in this study, in accord with the value obtained by
Eenens \& Williams (1992) based on IR recombination lines. 
WCL stars are more problematic since recombination coefficients
for C\,{\sc iii} are not available. In addition, the fraction of 
recombined helium is not straightforward to assess in such studies, 
so that helium abundances could be underestimated.

Oxygen abundances were more difficult to constrain, as already discussed
by Hillier \& Miller (1999), with the principal diagnostic region spanning
$\lambda\lambda$2900--3500. Since this spectral region was absent for WR146,
we adopted C/O=4 by number, as predicted by current stellar 
evolutionary models. Note that our neon abundance determination in 
Sect.~\ref{sect7} is relatively insensitive to the precise oxygen content. 
Regarding silicon and iron, we  adopt solar abundances since all distances 
are $\le$2 kpc.

We now proceed to discuss individual stars in detail.


\subsection{HD\,156385 (WR\,90)}


The principal datasets for WR90 comprised HIRES IUE ultraviolet
spectroscopy and AAT/RGO spectrograph optical observations. Secondary
datasets were the blue CTIO spectroscopy from Torres-Dodgen \& Massey (1988)
and the 2.6-5$\mu$m ISO spectroscopy (longer wavelength data were
 of insufficient S/N to be used as stellar diagnostics). The combined,
UV-optical-IR flux calibrated dataset for WR90 allowed a 
well constrained reddening of $E_{\rm B-V}$=0.38 mag, in agreement with
the determination by Morris et al. (1993) (see Table~\ref{table2}).
A distance of 1.55 kpc is implied from our assumed WC7 $M_{v}$-calibration.

The difficulty in identifying the stellar continuum in the rich emission
line spectrum of WC stars makes rectification imprecise, as
emphasised by Hillier \& Miller (1998, 1999). 
Fig.~\ref{fig1} demonstrates the excellent agreement between the line 
and continuum distribution of the model spectrum (dotted lines) and 
observations (solid lines), and
includes the true theoretical continuum distribution (dashed lines).
The number of line features that are poorly reproduced is small, and
includes C\,{\sc iii} $\lambda$5696 (too weak), $\lambda$9710 (too strong), 
plus C\,{\sc iv} $\lambda$1548--51, Si\,{\sc iv} $\lambda$1393--1402 
(both too weak because of Fe-absorption) and 
O\,{\sc vi} $\lambda\lambda$3811--34 (too weak). The use of 
absolute fluxes appears to give poor results in the line strengths around 
C\,{\sc iii} $\lambda$2297 since the de-reddened and theoretical 
continuum do not exactly match.

Our analysis reveals stellar parameters 
of $T_{\ast}$=71,000K, log($L/L_{\odot}$)=5.5 and
$\dot{M}/\sqrt{f}$=8$\times$10$^{-5}$ \msunyr. Adopting a volume
filling factor of $f$=0.1 indicates
log $\dot{M}/(M_{\odot}$ yr$^{-1})=-$4.6
and a wind performance number of $\sim$8.
Use of C\,{\sc iv} $\lambda$5471/He\,{\sc ii} $\lambda$5412 revealed
C/He=0.25$\pm$0.05 by number. An oxygen abundance of O/He=0.03$\pm$0.01 by number
was obtained by matching O\,{\sc iv} $\lambda\lambda$2916--26, 
$\lambda\lambda$3063--72, and $\lambda$3560--63.

\subsection{HD\,192103 (WR\,135)} \label{model_wr135}

\begin{figure}
\vspace{18.5cm}
\includegraphics{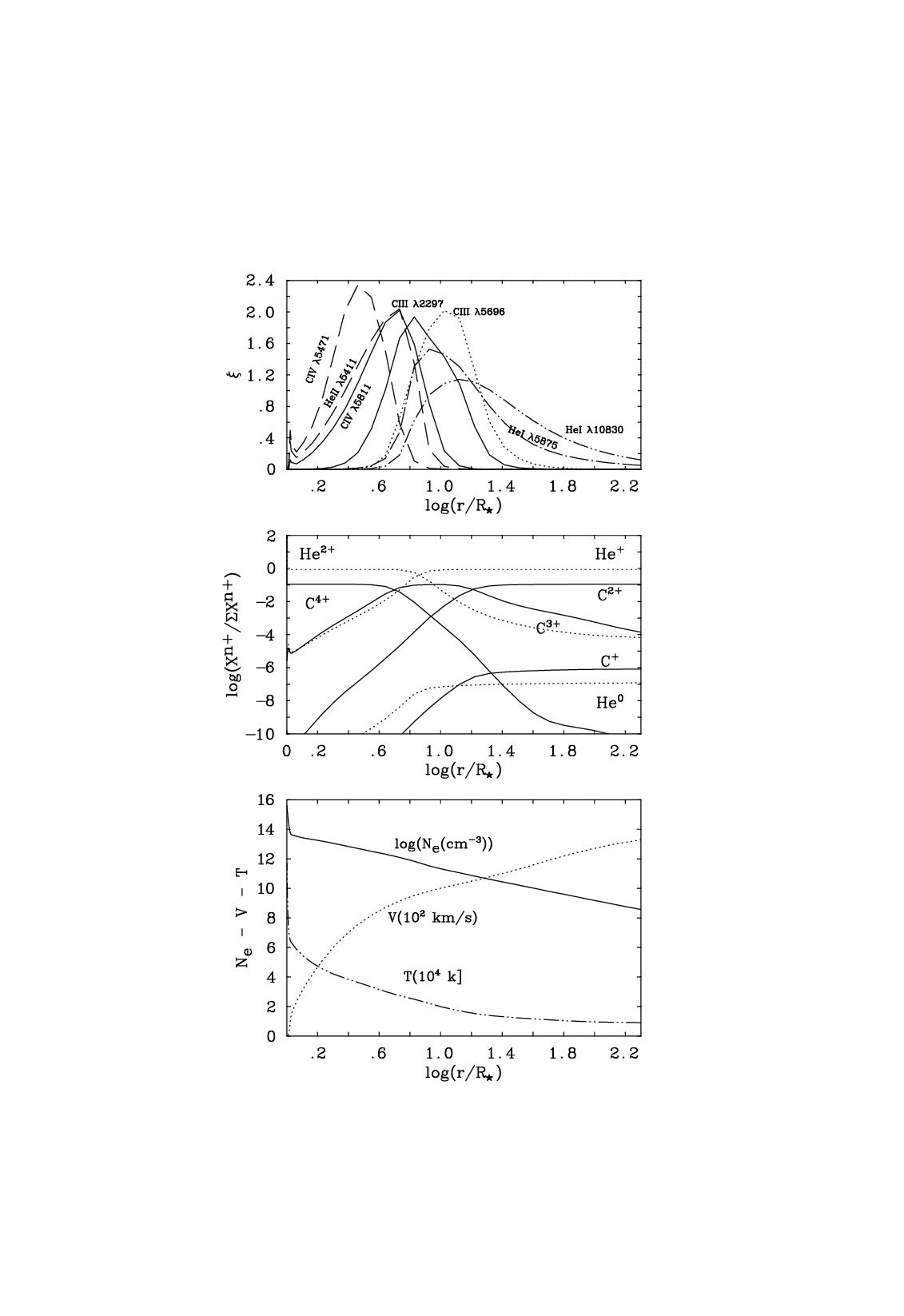}
\caption{Model predictions for the WC8 star WR135, analysed 
in Section~\ref{model_wr135}
{\bf (Top)}: Line formation regions for a selection of lines. 
The parameter $\xi$ is 
related to the observed flux emitted in the corresponding line, 
and defined as in 
Hillier (1987). The integral of $\xi$ over d$\log (r/R_{\ast})$ is
proportional to the equivalent width. {\bf (Middle)}: Wind ionization 
stratification for carbon (solid) and helium (dotted). 
Note that helium 
does not recombine to its neutral state in the outer wind.
{\bf (Bottom)}: Radial dependence of electron temperature (kK, dot-dash), 
density (cm$^{-3}$, solid), and wind velocity (km\,s$^{-1}$, dotted)}
\label{fig_wind}
\end{figure}

A substantial observational dataset was available for the WR135
analysis, particularly high quality UV (HIRES IUE), optical (INT), near-IR
(INT, UKIRT) and mid-IR (ISO) spectrophotometry. An interstellar reddening of
$E_{\rm B-V}$=0.37 was obtained for WR135, in accord with Smith et al. (1990)
(see Table~\ref{table2}), but 0.18 mag lower than the more 
recent determination of Morris et al. (1993). Membership of Cyg~OB3 implied an absolute 
visual magnitude of $M_{v}=-$4.3 mag.

\begin{figure*}
\vspace{23cm}
\includegraphics{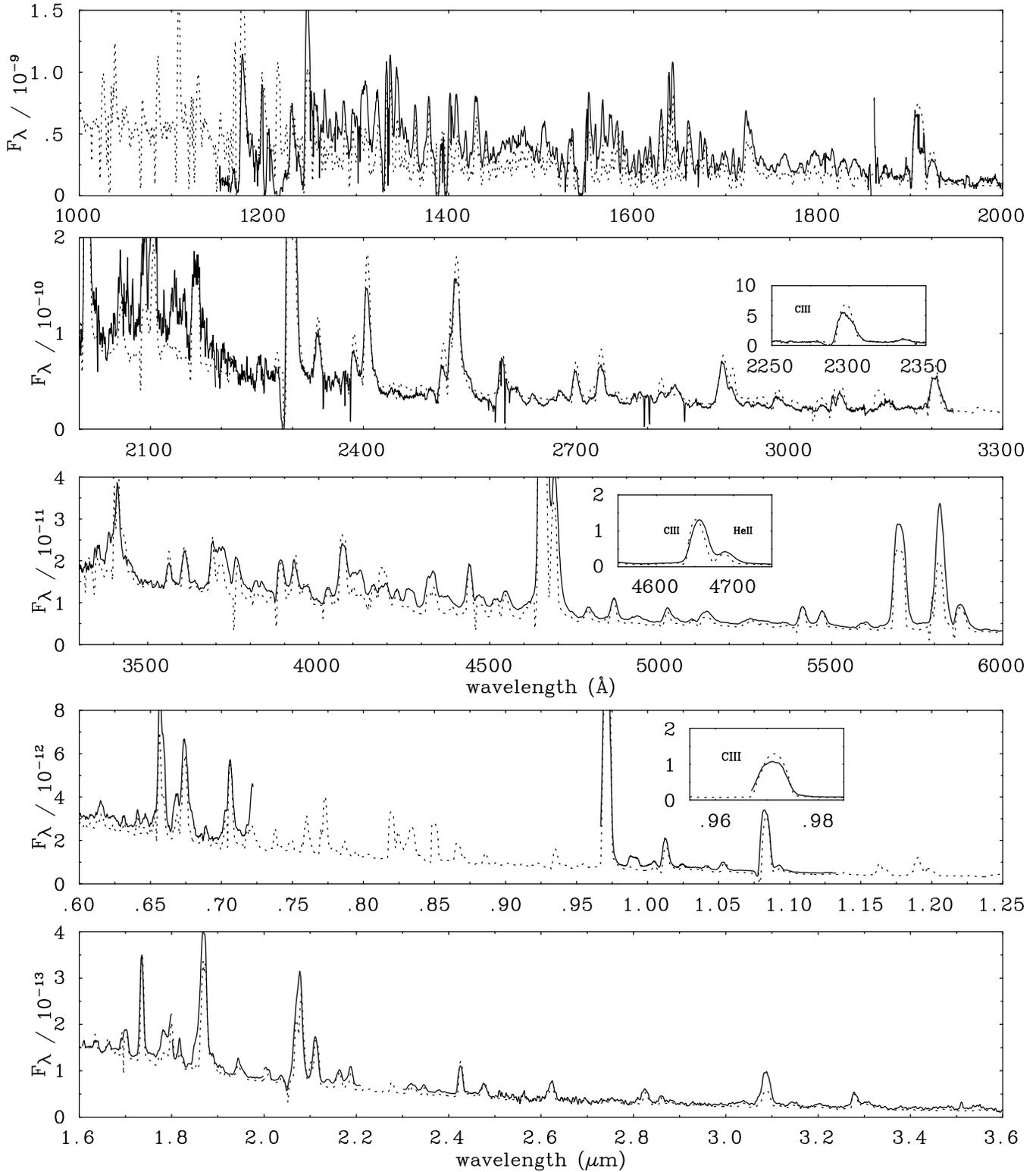}
\caption{Comparison between the de-reddened (E$_{B-V}$=0.37 and R=3.1)
spectrophotometry of HD\,192103 = WR135 (WC8) obtained from
 IUE, INT, UKIRT and ISO (solid lines) and
line-blanketed, clumped model predictions (dotted lines). 
Stellar parameters are $\mdot$=1.3$\times$10$^{-5}$\msunyr, 
$T_{\star}$=$63~kK$, log\,$L/L_{\odot}=5.20$, C/He=0.13 and 
O/He=0.03 by number. Flux units are in ergs/cm$^{2}$/s/\AA.}
\label{fig2}
\end{figure*}

Dereddened spectroscopy of WR135 (solid lines) is compared to our final 
synthetic model (dotted lines) in
Figure~\ref{fig2}. Agreement is overall excellent, even for  the
forest of iron lines in the UV. The relatively low wind velocity of  WR135
permits a greater number of individual diagnostics to be selected,  including
He\,{\sc ii} $\lambda$4686, He\,{\sc i} $\lambda$5876, so that the derived
properties can be treated with confidence. Indeed, 
C\,{\sc iii} $\lambda$5696 and C\,{\sc iv} $\lambda$5804 are fairly
well matched in this case. The strong C\,{\sc iii} spectral features at 
$\lambda$2297 and $\lambda$9710 are predicted to be 
20--30\% too strong. Other notable model deficiencies include 
underestimating the strength of 
C\,{\sc ii} emission at $\lambda$4267, $\lambda$9900, and O\,{\sc vi} 
$\lambda\lambda$3811--34, with O\,{\sc iii} $\lambda$3130 too strong. 
As with WR90, the use of absolute fluxes, appears 
to  give poor results in the line strengths around He\,{\sc ii} 
$\lambda$2530 since the de-reddened and theoretical continuum do not 
exactly match. This discrepancy simply reflects the inadequacy of the
adopted reddening law, rather than a fundamental flaw with current models.

We obtain the following stellar parameters: T$_{\ast}$=63\,kK, 
log($L/L_{\odot}$)=5.2 and
$\dot{M}/\sqrt{f}$=3.8$\times$10$^{-5}$\msunyr. Adopting a volume
filling factor of $f$=0.1 indicates log $\dot{M}/(M_{\odot}$ yr$^{-1})=-$4.9
and a wind performance number of $\sim$8. Use of C\,{\sc iv} 
$\lambda$5471/He\,{\sc
ii} $\lambda$5412 revealed C/He=0.13$\pm$0.03 by number. Once 
again, the oxygen abundance is poorly constrained, although 
O/He=0.03$\pm$0.01 provides a reasonable
match to O\,{\sc iv} $\lambda\lambda$2916--26 and 
$\lambda\lambda$3063--72, with O\,{\sc iii} $\lambda$3127 too strong.

In Fig.~\ref{fig_wind}, we 
present the wind structure of the final WR135 synthetic model,
including line formation regions, ionization balance etc.
Similar relations are shown for WR111 (WC5) in Hillier \& Miller (1999). 

\begin{table*}
\caption[]{Derived stellar properties for programme WC stars, including
results obtained by De Marco et al. (2000) for WR11 and 
Hillier \& Miller (1999) for WR111 for comparison}
\label{table3}
\begin{center}
\begin{tabular}{l@{\hspace{2mm}}c@{\hspace{2mm}}c@{\hspace{2mm}}c@{\hspace{2mm}}c@{\hspace{2mm}}
c@{\hspace{2mm}}c@{\hspace{2mm}}c@{\hspace{2mm}}c@{\hspace{2mm}}c@{\hspace{2mm}}c@{\hspace{2mm}}
c@{\hspace{2mm}}c@{\hspace{2mm}}c}
\hline
WR& Spectral& $T_{\ast}$ & $T_{2/3}$ &log (\lsun)& $\vinfty$&log $\mdot$&$\dot{M}/\sqrt{f}$
&C/He&O/He &log $Q_{{\rm H}^0}$ &log$Q_{{\rm He}^0}$ & M$_{v}$(WC)\\
  & Type & kK & kK &   &  \kms  &\msunyr &10$^{-5}$\msunyr& $\#$ & $\#$ & s$^{-1}$& s$^{-1}$ & mag\\
\hline
11   &WC8+O7.5\,III & 57 & 51 & 5.0  &  1550&$-$5.1& 2.9 & 0.15  & 0.03 & 48.8  & 47.8  & $-$3.7 \\
%
%
90   &WC7    & 71 & 29 & 5.5  &  2045&$-$4.6& 8.0 & 0.25  & 0.03 & 49.3   & 48.7  & $-$4.7 \\
111  &WC5    & 91 & 30 & 5.3  &  2300&$-$4.8& 4.7 & 0.4   & 0.10 & 49.2   & 48.5  & $-$4.2 \\
%
%
135  &WC8    & 63 & 27 & 5.2  &  1400&$-$4.9& 3.8 & 0.13  & 0.03 & 49.1  & 48.3  & $-$4.3 \\
%
%
146  &WC5+O8& 57 &33 & 5.7  &  2700&$-$4.5&10.5 & 0.08  & 0.02 & 49.6  & 48.7  & $-$5.3 \\
\hline
\end{tabular}
\end{center}
\end{table*}

\subsection{WR\,146}

Our principal observational datasets for WR146 are identical to those
used in Paper~I, namely INT (optical), UKIRT (near-IR) and ISO (2.6-5$\mu$m
because of the  low S/N at longer wavelengths). The difference 
in our approach is to derive stellar and chemical properties solely 
from spectral
synthesis, rather than recombination theory and independent modelling of the
continuum. 
As discussed in Sect.~\ref{wr146_dist}, we follow the distance estimate
of 1.4kpc from Dougherty et al. (2000), in the light of new observations from
Niemela et al. (1998).

\begin{figure*}
\vspace{14cm}
\includegraphics{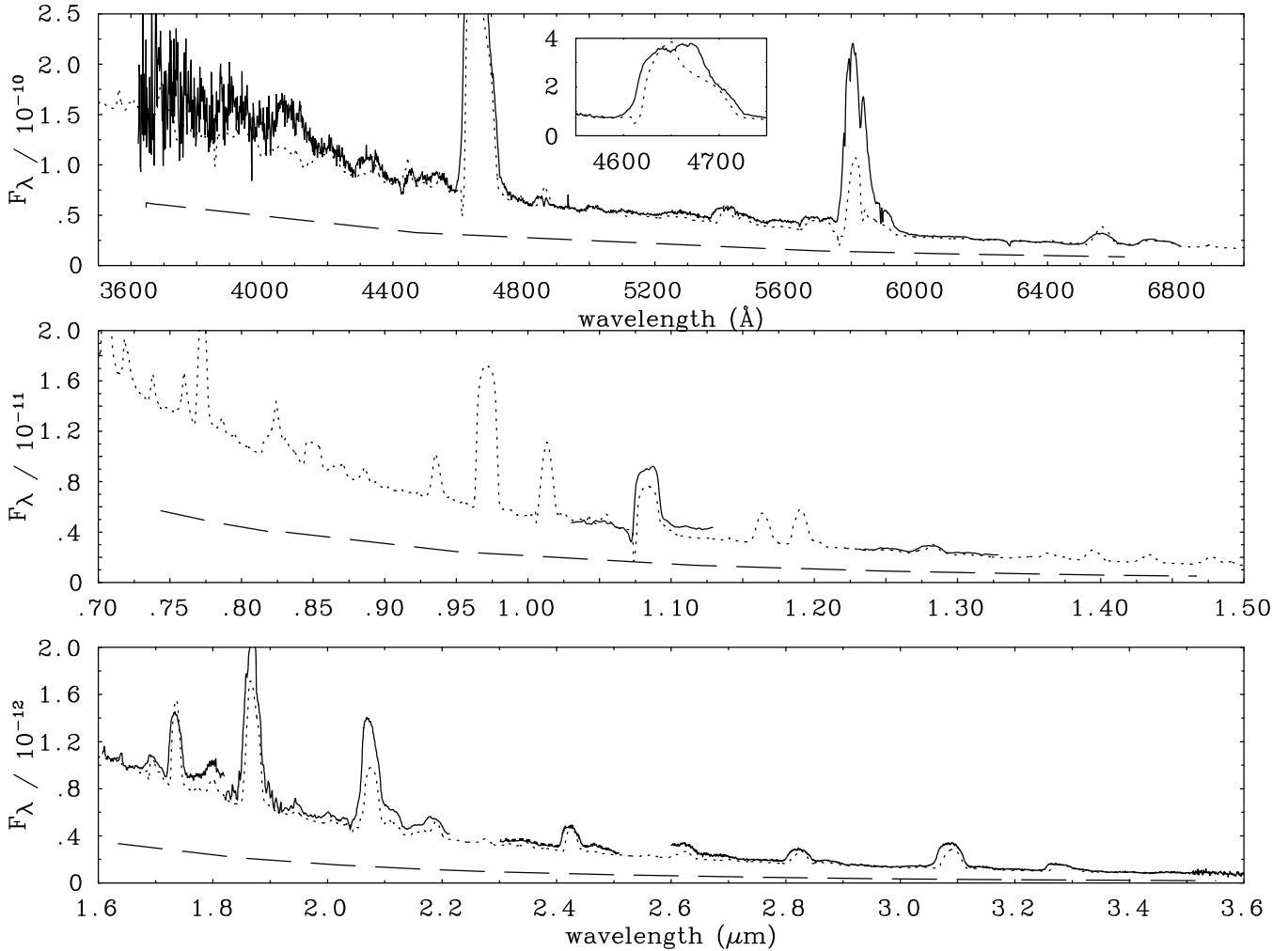}
\caption{Comparison between the de-reddened ($E_{\rm B-V}$=2.87 mag, R=2.9)
spectrophotometry of WR\,146 (WC5+OB) obtained from INT, UKIRT and ISO (solid
lines) and 
line-blanketed, clumped model predictions (dotted lines). The 
contribution of the late O companion is taken into consideration,
using model calculations by Crowther \& Bohannan (1997) -- its
continuum distribution is indicated with dashed-lines. 
WC stellar parameters are $\mdot$=3.3$\times$10$^{-5}$\msunyr, 
$T_{\star}$=$57~kK$, log\,$L/L_{\odot}=5.7$, C/He=0.08 and 
O/He=0.02 by number. Flux units are in ergs/cm$^{2}$/s/\AA.}
\label{fig3}
\end{figure*}

We have included the spectral energy distribution of a O8
supergiant in our
synthesis, using that for HD\,151804 (O8\,If) derived by 
Crowther \& Bohannan (1997). The IR free-free excess of this model
is somewhat greater than equivalent temperature Kurucz (1991) models 
for which the lowest gravity available is log~$g$=4.
Because of the contamination from the late O supergiant, the line spectrum 
of the WC component of WR146 is relatively weak, as shown in Fig.~\ref{fig3}. 
The flux level of the O companion, the contribution of which
declines with increasing wavelength, is illustrated as a dashed line in 
Fig.~\ref{fig3}. Comparison between de-reddened observations and
our synthetic spectrum is excellent, with the notable exception of 
C\,{\sc iv} $\lambda$5804. 

Our final stellar parameters for
the WC5 star were, T$_{\ast}$=$57~kK$, log\,$L/L_{\odot}=5.7$, and 
$\dot{M}/\sqrt{f}$=1$\times$10$^{-4}$\msunyr. A filling factor of $f$=0.1
reproduced the red wing of C\,{\sc iii-iv} $\lambda$4650--He\,{\sc ii}
$\lambda$4686. The high wind velocity of WR146 meant that C\,{\sc iv} 
$\lambda$5471/He\,{\sc ii} $\lambda$5412 were blended.
Nevertheless, we were able to constrain the carbon content, deriving 
C/He=0.08$\pm$0.02 by number, a factor of two times lower than that 
estimated in Paper~I from recombination line analysis, but now in
accord with Eenens \& Williams (1992). Oxygen is extremely difficult
to measure in WR146 since the usual near-UV diagnostics are unavailable, 
so that O/He=0.02$\pm$0.01 is adopted.

Finally, should the absolute magnitude of WR146 be more typical
of other WCE stars, namely $M_{v}$=$-$3.7 mag (Smith et al. 1990),
what would be the effect on the derived stellar parameters? 
In this case, our reddening towards WR146 would imply a distance of 
660 pc, such that the companion star would be extremely faint 
$M_{v}$=$-$3.5 mag, typical of an early B dwarf. The stellar properties
of WR146 would be unchanged, except that log\,$L/L_{\odot}=5.0$, and 
$\dot{M}/\sqrt{f}$=3.3$\times$10$^{-5}$\msunyr. In this case it 
would be difficult to reconcile these properties with the HST/radio 
observations of WR146 (Dougherty et al. 1996; Niemela et al. 1998).

%

\section{Summary of spectroscopic results}\label{sect6}

In this section we discuss the results of our quantitative 
analyses, and make comparisons with previous studies.

\subsection{Stellar parameters of WC stars}

In Table~\ref{table3}, we provide a summary of the derived properties 
of our programme WC stars, including results for WR11 and WR111
from De Marco et al. (2000) and Hillier \& Miller (1999),
obtained using identical techniques. Despite our sample 
spanning WC5 to WC8, there 
 is no obvious trend between spectral type and stellar temperature. Indeed,
the WC5 component of WR146 is found to have the lowest stellar temperature,
although this is certainly a most unusual WCE star! The 
subtle behaviour of C\,{\sc iii}\,5696\AA\  and C\,{\sc iv}\,5808\AA\
from our modelling indicates that, unfortunately, we are unable to 
use these as probes of wind ionization for WC5--8 stars (while we cannot obtain
a simultaneous fit to both lines, a relatively small
change in parameters can lead to a fit of either line).

Similarly, although there is considerable spread in carbon abundances, 
WR146 exhibits the lowest carbon mass
fraction, in contrast to the predicted increase in C/He ratio at
earlier spectral type. Koesterke \& Hamann (1995) derived a broad range
of C/He ratios at each spectral type for WC5--8 stars.

How do the present line blanketed, clumped results compare with previous
studies? Unfortunately, WR135 is the sole programme star that has been 
the subject of quantitative studies in the past. Eenens \& Williams (1992)
derived elemental abundances from IR recombination lines, estimating
C/He=0.12 by number, in excellent agreement with our determination of 0.13.
%
%
Howarth \& Schmutz (1992)
used a pure helium non-LTE model analysis to investigate WR135 based solely on
1$\mu$m spectroscopy. Since pure helium models are expected to be 
inadequate for WC analyses, Koesterke \& Hamann (1995) considered
both carbon and helium in their study of WC stars, including WR135.

In Table~\ref{tablekoesterke} we compare results from our present analysis
of WR135 with these previous studies, scaling their results to our absolute
visual magnitude. Wind velocities and $\dot{M}/\sqrt{f}$ are
in relatively good agreement, while the effect of including
carbon and blanketing has a dramatic effect on the stellar temperatures
and luminosities, such that the luminosity derived here is a factor of two
times higher than Howarth \& Schmutz (1992), who adopted a stellar 
temperature of 35,000K.

Stellar parameters are in much better agreement with Koesterke 
\& Hamann (1995). They compared specific line strengths with model grids 
at fixed C/He ratio, and chose simple model atoms of He\,{\sc i-ii} and
C\,{\sc ii-iv}. Although
spectral comparisons between model predictions and observations were not
presented by Koesterke  \& Hamann (1995), the fit quality was judged to be
poor, with large discrepancies for the C\,{\sc iii} $\lambda$5696 and
$\lambda$6740 lines. Therefore, we have greater confidence in our 
results since detailed UV, optical and IR synthetic spectrophotometry compare 
favourably with observations. 

From Table~\ref{tablekoesterke}, blanketing and clumping conspire to 
revise the wind performance number, $\dot{M}/v_{\infty}/(L/c)$,
from 30 in the study of Koesterke \& Hamann (1995) to just 5 for WR135! 
For our entire sample, performance numbers are $\le$10, with
previous studies indicating values of up to 100 (Howarth 
\& Schmutz 1992; Koesterke \& Hamann 1995).

\subsection{Evolutionary status}

The wide range in stellar luminosity of our sample of WC
stars, $L/L_{\odot}$=10$^{5}$ to 10$^{5.7}$, implies 
a considerable range in current stellar masses. Following the 
mass-luminosity relation for hydrogen-free WR stars of 
Schaerer \& Maeder (1992), present masses of 
9$M_{\odot}$ (WR135), 13$M_{\odot}$ (WR90) and 19$M_{\odot}$ (WR146)
are implied. In contrast, initial mass estimates are much more
dependent on specific evolutionary tracks.

All models predict stars to pass through the WC phase at ages
of between 2.7--4.5Myr. For initial 40 and 60$M_{\odot}$ models,
stellar masses during the early WC phase are predicted to 
be $\sim$14$M_{\odot}$
and $\sim$24$M_{\odot}$, respectively. At higher initial mass, the 
situation is extremely model dependent. For example, an evolutionary model 
with initial mass 120$M_{\odot}$ will produce a 60$M_{\odot}$ WC star using 
standard de Jager et al. (1988) mass-loss rates. Alternatively,
a 10$M_{\odot}$ WC star will result based on (2$\times$) enhanced 
mass-loss for the post main-sequence and WNL phases 
(Schaller et al. 1992), or only a 4$M_{\odot}$ WC star for 
(2$\times$) enhanced mass-loss during the entire stellar evolution 
(Meynet et al. 1994). Therefore, the present masses and ages of our
programme WC stars are well constrained, but initial masses 
are not. 

\begin{figure}
\vspace{8cm}
\includegraphics{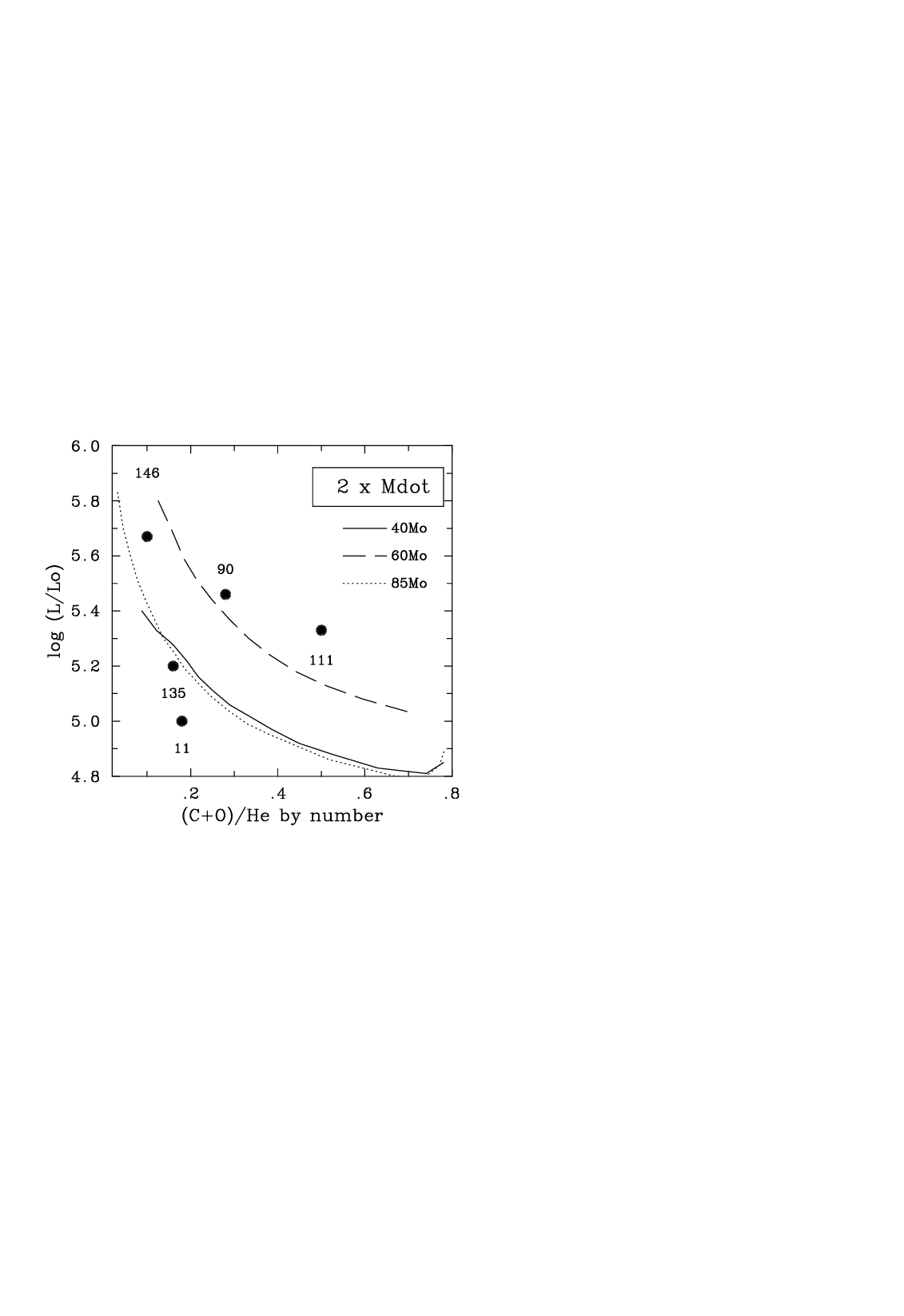}
\caption{Comparison of measured (C+O)/He ratios versus
luminosities for our programme stars,  including WR111 (WC5)
from Hillier \& Miller (1999), with theoretical expectations 
from the solar metallicity evolutionary tracks 
assuming enhanced (2$\times$) mass-loss (Meynet et al. 1994)}
\label{fig5a}
\end{figure}

In order to attempt estimates of initial masses, we compare our 
derived (C+O)/He ratios versus stellar 
luminosities with evolutionary predictions in Fig.~\ref{fig5a}. 
Predictions for 40--120$M_{\odot}$ initial models assume
(2$\times$) enhanced mass-loss relative to 
de Jager et al. (1988) during the entire stellar evolution
(Meynet et al. 1994). These models favour 60$M_{\odot}$ for WR146 and WR90
and 40$M_{\odot}$ {\it or} 85$M_{\odot}$ for WR135. Consequently,
multiple initial mass estimates may result for individual stars.
Nevertheless, we currently favour 40$M_{\odot}$ for WR135 and 
60$M_{\odot}$ for WR146 and WR90.

Specific comparisons between  the stellar properties of WR90, 
WR135 and WR146 and Meynet et al. (1994) evolutionary predictions
are made in Table~\ref{evolution}. Both the initial 60$M_{\odot}$
and 85$M_{\odot}$ models adopt mass-loss rates that exceed observations
by large factors during early WC stages. This is because evolutionary
models for hydrogen-free WR stars adopt mass-loss rates that are
solely functions of mass (Langer 1989). Clearly, future evolutionary 
models should take allowance for more appropriate WR mass-loss rates.

\begin{table}
\caption[]{Comparison of our derived stellar parameters for 
WR\,135 with Howarth \& Schmutz (1992, HS92) and 
Koesterke \& Hamann (1995, KH95). We have adjusted their parameters
to our assumed visual absolute magnitude ($M_{v}=-$4.3 mag)}
\label{tablekoesterke}
\begin{center}
\begin{tabular}{l@{\hspace{1mm}}l@{\hspace{-3mm}}c
@{\hspace{1mm}}c@{\hspace{1mm}}c@{\hspace{1mm}}c@{\hspace{1mm}}c}
\hline
Ref. &Model &    log(\lsun)& $T_{\ast}$  &$\vinfty$  &   $\mdot$/$\sqrt{f}$  &C/He     \\
     &      &        &    kK        & \kms      &   \msunyr    & \\
\hline
HS92     & He       & 4.9 &(35) & 1500  & 5$\times 10^{-5}$ & --    \\ 
KH95     & He+C     &5.1 & 76 &  1300  & 6$\times 10^{-5}$ & 0.14  \\ 
This work& He+C+O+Fe&5.2 & 63 & 1400  & 4$\times 10^{-5}$ & 0.13  \\ 
\hline
\end{tabular}
\end{center}
\end{table}

\subsection{Radio fluxes}\label{radio}

We now compare predicted radio fluxes from our spectroscopic mass-loss 
rates. Hillier \& Miller (1999) provide a formulation 
based on Eqn 9 in Wright \& Barlow (1975), allowing 
for the filling factor according to Abbott et al. (1981).

In all cases, the dominant ionization at the outer boundary of our models,
$N_e \sim 10 ^{8}$ cm$^{-3}$ or 200$R_{\ast}$ is He$^{+}$, C$^{2+}$ and 
O$^{2+}$ (see Fig.~\ref{fig_wind}). Since the radio 
emitting region lies at lower densities, we 
also consider cases in which C$^{+}$ and/or O$^{+}$ are dominant.
Outer wind electron temperatures are typically 8000K (Fig.~\ref{fig_wind}), 
except for WR146 for which the lower cooling produced by less
metals implies 9000K. 

\begin{table}
\caption[]{Comparison between our derived stellar parameters for 
WC stars and evolutionary predictions from Meynet et al. (1994)}
\label{evolution}
\begin{center}
\begin{tabular}{l@{\hspace{1mm}}l@{\hspace{1mm}}c@{\hspace{1mm}}c@{\hspace{1mm}}c
@{\hspace{1mm}}c@{\hspace{1mm}}c@{\hspace{1mm}}c}
\hline
WR  & $M$ & $\tau$ &$T_{\ast}$ & log(\lsun)& $\mdot$  &C/He & O/He   \\
  &$M_{\odot}$&Myr &kK         &           &  \msunyr &     &        \\
\hline
90   &  &  & 71        &5.5        &2.5$\times 10^{-5}$& 0.25 & 0.03 \\
135  &  &  & 63        &5.2        &1.2$\times 10^{-5}$& 0.13 & 0.03 \\
146  &  &  & 57        &5.7        &3.3$\times 10^{-5}$& 0.08 & 0.02 \\
\hline
     & \multicolumn{5}{c}{---  $M_{\rm init}$=40 M$_{\odot}$ ---} \\
     &13&4.56&126      &5.4        &5.5$\times 10^{-5}$& 0.08 & 0.004 \\
     & 8&4.72&122      &5.1        &1.8$\times 10^{-5}$& 0.25 & 0.04  \\
     & \multicolumn{5}{c}{---  $M_{\rm init}$=60 M$_{\odot}$ ---} \\
     &23&3.57&136      &5.8       &24.6$\times 10^{-5}$& 0.12 & 0.01 \\
     &12&3.67&131      &5.4       & 4.8$\times 10^{-5}$& 0.25 & 0.04  \\
     & \multicolumn{5}{c}{---  $M_{\rm init}$=85 M$_{\odot}$ ---} \\
     &15&3.10&128      &5.5        &9.1$\times 10^{-5}$& 0.07 & 0.004 \\
     & 8&3.30&120      &5.0        &1.7$\times 10^{-5}$& 0.25 & 0.04  \\
\hline
\end{tabular}
\end{center}
\end{table}

We compare predicted
radio fluxes with observed values taken from the literature in
Table~\ref{tab10}, in which a uniform UV/optical filling factor
of $f$=0.1 has been adopted throughout. 
Note that the quoted radio flux for WR146 refers solely
to the WC component (the southern mean emitted flux S$_{5}$ in Dougherty et 
al. 2000). We find that consistency is 
excellent for WR135 and WR146 for doubly ionized carbon and oxygen, 
while singly 
ionized carbon and oxygen are favoured for WR90. However,
C$^{+}$ is not predicted to be the dominant ionization stage in the outer
wind of these stars (Fig.~\ref{fig_wind}), so that 
the predicted radio flux for this star appears to be too high.

\begin{table}
\caption[]{Comparison between predicted and observed 6cm (4.9GHz)
radio fluxes in our programme stars, assuming $f$=0.1 for the 
UV/optical line forming region. The filling factor entry in the table
corresponds to the adopted value in the radio region, indicating the
change in that quantity required to recover the observed
flux at 4.9GHz. We have included WR11 in this table, since a neon abundance
determination for this star is to be carried out in Sect.~\ref{neon}.}
\label{tab10}
\begin{tabular}{r@{\hspace{2.5mm}}l@{\hspace{1.0mm}}l@{\hspace{1.0mm}}
c@{\hspace{0.5mm}}c@{\hspace{0.5mm}}c@{\hspace{1.5mm}}r
@{\hspace{1.5mm}}r}
\hline
WR          & \multicolumn{3}{c}{Ionization}& $T_{e}$ & $f$ &
\multicolumn{2}{c}{S$_{\nu}^{\rm 6cm}$ (mJy)} \\ 
             & \multicolumn{3}{c}{state}    & kK     &radio& Pred. & Obs.(Ref) \\                      
\hline\noalign{\smallskip}
11& He$^{+}$ &C$^{2+}$  & O$^{2+}$& 8 &0.10&  20.5  & \\ 
  & He$^{+}$ &C$^{2+}$  & O$^{2+}$& 8 &0.05&  32.5  & 32.2 (a)     \\ 
   & He$^{+}$ &C$^{2+}$  & O$^{+}$ & 8 &0.10& 20.2  &  \\ 
   & He$^{+}$ &C$^{+}$   & O$^{+}$ & 8 &0.10& 15.1 &     \\ 
\noalign{\smallskip}
90 & He$^{+}$ & C$^{2+}$ &O$^{2+}$ & 8 &0.10&   1.72 & \\ 
   & He$^{+}$ & C$^{2+}$ &O$^{2+}$ & 8 &0.20&   1.08 &1.10 (b)         \\ 
   & He$^{+}$ &C$^{2+}$  & O$^{+}$ & 8 &0.10&   1.64 &         \\ 
   & He$^{+}$ &C$^{+}$   & O$^{+}$ & 8 &0.10&   1.10 &         \\ 
\noalign{\smallskip}
135& He$^{+}$ &C$^{2+}$  & O$^{2+}$& 8 &0.10&   0.54 &   0.60 (a) \\ 
   & He$^{+}$ &C$^{2+}$  & O$^{+}$ & 8 &0.10&   0.51 &        \\ 
   & He$^{+}$ &C$^{+}$   & O$^{+}$ & 8 &0.10&   0.40 &        \\ 
\noalign{\smallskip}
146& He$^{+}$ &C$^{2+}$  & O$^{2+}$& 9 &0.10&   2.04 &  2.00 (c) \\ 
   & He$^{+}$ &C$^{2+}$  & O$^{2+}$& 9 &0.20&   1.29 &           \\ 
   & He$^{+}$ &C$^{2+}$  & O$^{2+}$& 9 &0.30&   0.98 &           \\ 
   & He$^{+}$ &C$^{2+}$  & O$^{+}$ & 9 &0.10&   1.95 &        \\ 
   & He$^{+}$ &C$^{+}$   & O$^{+}$ & 9 &0.10&   1.65 &        \\ 
\noalign{\smallskip}\hline
\end{tabular}
\begin{flushleft}
a: Leitherer et al (1997), b: Abbott et al. (1986), c: Dougherty et al. (2000)
\end{flushleft}
\end{table}

It is possible that the filling factor in the radio and optical forming 
regions differs significantly. The predicted
radio flux of WR90 would be in excellent agreement with the
observed value if the radio filling factor was $\sim$0.2,
while that of WR11 requires $\sim$0.05. Hillier \& Miller (1999) 
found a similar discrepancy in their study of WR111. Note also 
that there is  observational evidence that WR90 may not be 
single, since it is a non--thermal emitter 
(Leitherer et al 1997; Chapman et al 1999).

We also include calculations for WR11 in Table~\ref{tab10}, since we attempt 
to re-derive its neon abundance determination in Section~\ref{sect7}.
As for the other stars in our sample, possible variations in 
filling factor between the inner and outer wind are important. 

\section{Neon and sulphur abundances in WC stars}\label{sect7}

We are now in a position to determine neon abundances in our
programme WC stars based on ISO spectroscopy, supplemented by
sulphur determinations for WR11.
In this section we first provide a revised
formulation for the determination of neon in a clumped medium, following
BRA88, and subsequently provide measurements for each star.

\subsection{Ionic abundances from fine-structure lines in an 
inhomogeneous wind}

We have re-derived the numerical and analytical forms for the determination
of ionic abundances in clumped winds from BRA88.
We consider a fine-structure line from ion $i$, with transition energy
$h \nu_{ul}$. If $D$ is the distance to the star and $I_{ul}$ is the
observed line flux, then 
\begin{equation}
4 \pi D^{2} I_{ul} = \int_{0}^{\infty} n_{u} A_{ul} h \nu_{ul} 
4 \pi r^{2}fdr \hspace*{1cm} {\rm erg~s}^{-1} \label{bra5}
\end{equation}
where $A_{ul}$ is the line transition probability
and we have introduced the filling factor $f$ into their formulation.
$n_{u}$ represents the density of ions in the upper level, and can be written
as, 
\begin{equation}
  n_{u} = f_{u} n_{i}            \hspace*{1cm} {\rm cm}^{-3}   \label{bra6}
\end{equation}
where $n_{i}$ is the species ion density and $f_{u}$ is the fractional population
of the upper level. Upper level populations,
$f_{u}$, were determined for each ion by solving the equations of statistical 
equilibrium using {\sc equib} (Adams \& Howarth, priv. comm) for $\ge$30 electron
densities in the range 10$^{0}$ to 10$^{12}$ cm$^{-3}$, and 9 electron 
temperatures covering 5k to 14kK. Thus,
\begin{equation}
n_{u} = \frac{f_{u} \gamma_{i} A}{f r^{2}} \hspace*{1cm} {\rm cm}^{-3}   \label{bra7}
\end{equation}
where $\gamma_i$ is the fraction of all ions represented by ion species $i$
\begin{equation}
 \gamma_{i} = \frac{n_{i}}{\sum_{j}n_{j}} \label{bra8}
\end{equation}
and $A$ is the mass loss parameter (Eqn 8 from BRA88).
Combining  equations~\ref{bra5}--\ref{bra7}, the filling factor term cancels
out, leaving
\begin{equation}
I_{ul} = \frac{\gamma_{i}}{D^{2}} A_{ul} h \nu_{ul} A
\int_{0}^{\infty}f_{u}(r, f, T)dr 
\hspace*{1cm} {\rm erg~cm}^{-2} {\rm s}^{-1}\label{bra10}
\end{equation}
We deviate from BRA88  by carrying out the integral in 
density, rather than radial space. Since $r(N_{e}$) is a bijection, we 
can modify this integral so that the dependency of $f_{u}$ on $A$ and $f$
(and hence $\dot{M}$) is removed from the integral term. 
Equation~\ref{bra10} can then be modified 
to:
\begin{equation}
\int_{0}^{\infty}f_{u}(r,f, T)dr =  
\sqrt{\frac{\gamma_{e} A}{4f}} \int_{0}^{\infty}
\frac{f_{u}(N_{e},T)}{N_{e}^{1.5}}d N_{e}  \label{bra11}
\end{equation}
For a more straightforward numerical computation, Equation~\ref{bra11} 
is slightly adjusted:
\begin{equation}
\int_{0}^{\infty}\frac{f_{u}(N_{e},T)}{N_{e}^{1.5}}d N_{e} = ln(10)
\int_{0}^{\infty}\frac{f_{u}(N_{e},T)}{\sqrt{N_{e}}}d log(N_{e}) \label{bra12}
\end{equation}
The effect of this modification leads to a significant improvement over BRA88,
who used an average value for each region $\Delta$r. 
since the line formation region is a very sensitive function
of f$_{u}$. Also, the
dependence of the integral on $\mdot$ is removed.
Consequently,
BRA88 overestimated the integral term, producing a lower
elemental abundance. Our final numerical expression for the ion number fraction
$\gamma_i$, is (cgs units):
\begin{equation}
 \gamma_{i}=\frac{(4 \pi \mu m_{H} \vinfty)^{1.5}}{ln(10)f^{0.25}}
\left(\frac{\sqrt{f}}{\mdot}\right)^{1.5}
\frac{1}{F_{u}(T)}\frac{2D^{2}I_{ul}}{\sqrt{\gamma_{e}}A_{ul}h\nu_{ul}} \label{bra13}
\end{equation}
with 
\begin{equation}
F_{u}(T) = \int_{0}^{\infty}\frac{f_{u}(N_{e},T)}{\sqrt{N_{e}}}d log(N_{e}).
\end{equation}
Provided radio and recombination processes are used to derive
$\mdot$/$\sqrt{f}$, elemental abundances are only weakly dependent on the
distance ($\propto$ D$^{-0.25}$) since mass-loss rates depend on $D^{1.5}$.
Therefore elemental abundances obtained with filling factors of $f$=0.1 
or 1.0 differ by a factor of 1.8. 

\begin{figure}
\vspace{10cm}
\includegraphics{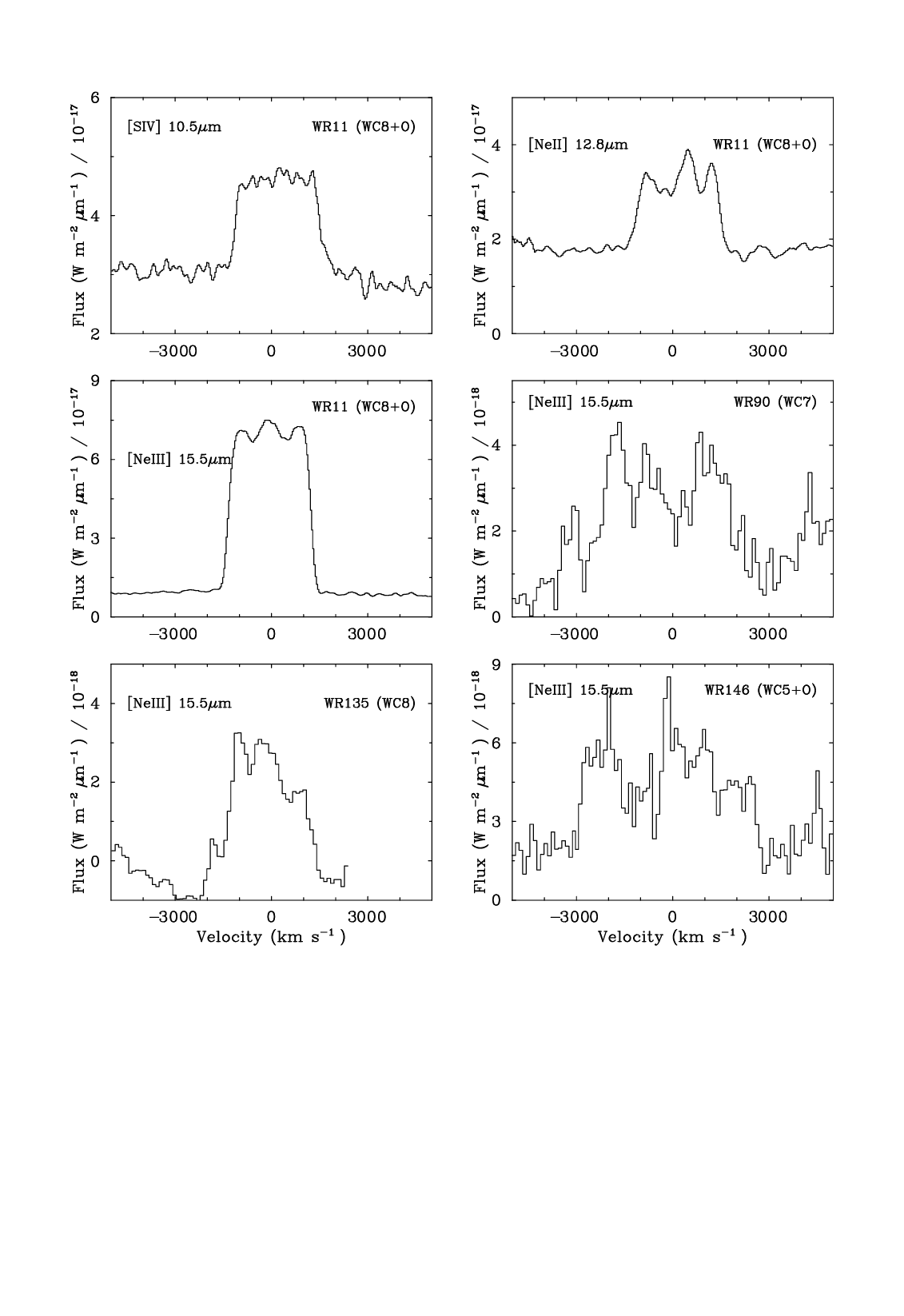}
\caption{ISO/SWS observations of fine structure [S\,{\sc iv}] 10.5$\mu$m,
[Ne\,{\sc ii}] 12.8$\mu$m and [Ne\,{\sc iii}] 15.5$\mu$m emission in 
WR11, WR90, WR135 and WR146.}
\label{fig4}
\end{figure}

In order to allow for the possibility of a
clumped medium, the analytical expression for the ion number fraction 
given by BRA88 (their equation A13) also needs to be multiplied by a 
factor of $\sqrt{f}$. We include determinations of $\gamma_{i}$ using
both the integral and analytical expressions in our subsequent calculations.
We find that the analytic expression is reliable to within $\sim$20\%.

\subsection{Neon abundances for WC stars}\label{neon}

We are now in a position to evaluate neon abundances in our programme
WC stars, using Equation~\ref{bra13}. Figure~\ref{fig4} shows 
ISO/SWS observations of [Ne\,{\sc iii}] 15.55$\mu$m emission for
each of our programme stars, together with 
[Ne\,{\sc ii}] 12.8$\mu$m in WR11. 
Emission line fluxes are listed in Table~\ref{table4a}.

Table~\ref{table4a} also includes atomic quantities used to derive 
ion fractions. Transition probabilities are taken from the NIST 
Atomic Spectra Database, while collision strengths are 
obtained from papers in the IRON Project series (see Table~\ref{table4a}
for references). ISO flux measurements for WR11 
compare closely with previous measurements 
from IRAS and ground-based observations (van der Hucht \& Olnon (1985); 
BRA88), as indicated in Table~\ref{table4a}.

In the following comparison, we include abundances 
derived from ISO mid-IR neon lines for WR11, using stellar parameters 
derived by De Marco et al. (2000).

Neon abundance determinations 
are sensitive to the ionization balance in the line forming region 
for the fine-structure lines, and is comparable to that of the radio
forming continuum, $N_{e}$$\sim$10$^{5}$ cm$^{-3}$. Ne$^+$,
with an ionization potential that is higher than C$^{+}$ and O$^{+}$,
is  observed solely in WR11 (WC8). Therefore, we have 
allowed for the possibility that the carbon and oxygen ionization 
balance are singly ionized for WC8 stars, with He$^{+}$ 
and doubly ionized carbon and oxygen otherwise. 
Table~\ref{table4} shows predicted ion abundances of 
Ne$^{2+}$ for each case, plus measured Ne$^{+}$ for WR11, with upper limits
otherwise. In all cases Ne/He$\sim$3--4$\times$10$^{-3}$ by number, significantly
greater than the expected cosmic value of Ne/He$\sim$5$\times 10^{-4}$ in 
the C and O-enriched WC environment.

It is possible that neon exists in (unseen) higher ionization stages, 
specifically Ne$^{3+}$. However, the ionization potential for Ne$^{3+}$,
97eV, is significantly higher than C$^{3+}$ (64eV) and O$^{3+}$ (77eV) 
which are not expected to be present in the outer winds of WC stars 
(Fig.~\ref{fig_wind}). 

In summary, measurement of fine-structure lines of neon from ISO/SWS observations
reveal Ne/He=0.003--0.004 (Ne/C$\sim$0.02), a factor of 6--8 times higher than cosmic 
abundances of Ne/He=0.0005 for the carbon rich winds of WC stars, supported also by
sulphur abundance determinations. However, what additional sources of uncertainty 
are there in our Ne/He determinations, and why do our results for 
WR11 differ from BRA88, who obtained Ne/He=0.001?

Although the ISO-SWS neon line fluxes are in good agreement with the 
ground-based and IRAS neon line fluxes from BRA88 (Morris et al. 1998), we 
derive an elemental abundance that is {\it three} times higher. The 
source of this discrepancy is due to a different distance to WR11, and 
the use of a non-clumped mass-loss rate by BRA88. Morris et al. (2000)
have recently emphasised the need for a reliable mass-loss rate estimate
in the determination of neon abundances. In addition, our derived 
neon abundance is in good agreement with Morris et 
al. (1998) who combined ISO/SWS neon 
fluxes with the {\sc hipparcos} distance to WR11, and the X-ray derived
mass-loss rate of Stevens et al. (1996).

The other major factor affecting neon abundances is clumping.
Our UV/optical analyses use filling factors of $f$=0.1, yet 
we cannot observationally constrain the filling factor to better 
than 0.05$\le f \le$0.25, indicating a further 20\% uncertainty in 
Ne/He.  Of greater importance, we assume identical filling factors
for the optical line forming region ($\approx$10$^{11}$ cm$^{-3}$) 
and the neon emitting region  ($\approx$10$^{5}$ cm$^{-3}$). 
Neon abundances would be increased by 40$\%$ for WR90,
decreased by 30$\%$ for WR11, 
with WR146 and WR135 unchanged, assuming that (i)  UV/optical mass-loss rates are 
fully consistent with radio fluxes (recall 
Sect.~\ref{radio}), and that (ii) volume filling factors in the neon 
emitting region are identical to those in the radio region. 

\subsection{Sulphur abundances for WR11}\label{sulphur}

In order to assess the reliability of our derived abundances, we have 
also calculated the sulphur abundance for WR11 based on ISO observations
of fine-structure [S\,{\sc iv}] 10.5$\mu$m  and [S\,{\sc iii}] 18.7$\mu$m lines
(Table~\ref{table4a}). Since sulphur is not enhanced by nucleosynthesis, abundances should 
correspond  to the cosmic value. The line forming region for the sulphur 
fine-structure lines peaks at $N_{e}$$\sim$10$^{4}$ cm$^{-3}$, somewhat lower than
the neon lines.

ISO spectroscopy of [S\,{\sc iv}] 10.5$\mu$m confirms the 
line flux measured from ground-based spectroscopy by BRA88, 
while stellar modelling anticipates weak contamination from
lines of He\,{\sc i} (12--8) 10.52$\mu$m and C\,{\sc iii} (20--17) 10.54$\mu$m. 
The stellar analysis of WR11 by De Marco et al. (2000) predicts a 15\% contribution
from these lines to the observed flux, which has been corrected accordingly.
For the first time in a Wolf-Rayet star, ISO reveals the presence of 
[S\,{\sc iii}] 18.7$\mu$m, blended with the stellar 
He\,{\sc i} (14--10) 18.62$\mu$m feature, resulting in a 30\% decrease in [S\,{\sc iii}]
flux. Use of an inappropriate mass-loss rate 
and distance for WR11 led BRA88 to suggest S$^{3+}$/He=2.5$\times 10^{-5}$, 
such that they were obliged to predict S$^{2+}>$S$^{3+}$, with an
expected high [S\,{\sc iii}] 18.7$\mu$m line flux that is not confirmed 
by ISO observations.

From Table~\ref{table4}, we find $\gamma_{{\rm S}^2+}=1.9\times 10^{-5}$ and
$\gamma_{{\rm S}^3+}= 3.1 \times 10^{-5}$, which imply S/He=6$\times 10 ^{-5}$ by number.
This is in good agreement with the cosmic value of 7.5$\times 10^{-5}$ for the 
C and He enriched environment of WR11. Therefore, our determinations imply
Ne/S=50 for WR11, a factor of eight times greater than the 
cosmic value of Ne/S$\sim$7. Since these lines are formed in similar regions of 
the stellar wind, Ne/S abundances are essentially independent of clumping, 
and reveal a degree of neon enrichment relative to sulphur that is 
equivalent to that derived earlier for He. 

\begin{table*}
\caption[]{Observed mid-IR fine structure line intensities (units of
10$^{-12}$ erg~cm$^{-2}$~s$^{-1}$), and adopted atomic parameters, including
statistical weights of the upper and lower levels, $\omega_u$ and $\omega_l$,
transition probability, $A_{ul}$, and collision strength $\Omega_{ul}$ at $T_e$=8000K.
Fluxes are obtained from ISO/SWS observations in all cases, except for literature
measurements for WR11, obtained from IRAS/LRS (van der Hucht \& Olnon 1985) and the UCL 
spectrometer at the AAT (BRA88).}
\label{table4a}
\begin{center}
\begin{tabular}{l@{\hspace{2mm}}c@{\hspace{2mm}}c@{\hspace{2mm}}c@{\hspace{2mm}}
c@{\hspace{2mm}}c@{\hspace{2mm}}c@{\hspace{2mm}}c@{\hspace{2mm}}c@{\hspace{2mm}}
l@{\hspace{0.5mm}}c@{\hspace{2mm}}c@{\hspace{2mm}}c@{\hspace{2mm}}c}
\hline
Ion            & Transition & $\lambda$&$\omega_u$ & $\omega_l$& $A_{ul}$& Ref
& $\Omega_{ul}$ & Ref& \multicolumn{2}{c}{WR11 (WC8+O)} & WR90 & WR135 & WR146 \\
               &            & $\mu$m   &           &           & s$^{-1}$ &    & 8000K         &    & ISO & IRAS/AAT & ISO (WC7) & ISO (WC8) & ISO (WC5+O)\\
\hline

{}[S\,{\sc iv}]  & $^2$P$^{\circ}_{3/2}$--$^2$P$^{\circ}_{1/2}$ & 10.51 & 4 & 2
& 7.70$\times$10$^{-3}$&a&8.47&c& 15$\pm$1$^{\ast}$ & 19$\pm$4 & --  & -- &  -- \\
{}[Ne\,{\sc ii}] & $^2$P$^{\circ}_{3/2}$--$^2$P$^{\circ}_{1/2}$ & 12.81 & 2 & 4 &
8.59$\times$10$^{-3}$&a&0.28&d& 18$\pm$1 & 17$\pm$3 & $\le$1 & $\le$0.5 & $\le$1 \\
{}[Ne\,{\sc iii}]& $^3$P$^{\circ}_{1}$--$^3$P$^{\circ}_{2}$     & 15.55 & 3 & 5
& 5.99$\times$10$^{-3}$&a&0.76&e& 82$\pm$1& 90$\pm$20 & 6.5$\pm$1&3.1$\pm$0.2&9.4$\pm$1\\
{}[S\,{\sc iii}]  & $^3$P$^{\circ}_{1}$--$^3$P$^{\circ}_{2}$ & 18.68 & 5 & 3
& 2.07$\times$10$^{-3}$&a&5.30&b& $<$1.8$^{\ast}$ & -- & --  & -- &  -- \\
\hline
\end{tabular}
\end{center}
(a) Naqvi (1951); (b) Galavis et al. (1995); (c) Saraph \& Storey (1999); 
(d) Saraph \& Tully (1994); (e) Butler \& Zeippen (1994); \newline
($\ast$) [S\,{\sc iii-iv}] line fluxes are shown prior to correction 
for the presence of He\,{\sc i} and C\,{\sc iii} stellar features (see text).
\end{table*}

\begin{table}
\caption[]{Neon and sulphur 
abundances derived for the programme stars, using the stellar parameters given in the
previous table. In all cases athe ionization balance is assumed to consist of
He$^{+}$, C$^{2+}$ and O$^{2+}$.
For each star, the first entry refers to the 
$\gamma_{i}$ derived from the (more reliable) integration method, with the 
second obtained from the analytical expression.}
\label{table4}
\begin{center}
\begin{tabular}{l@{\hspace{1.5mm}}l@{\hspace{1.5mm}}l@{\hspace{1.5mm}}
l@{\hspace{1.5mm}}l@{\hspace{1.5mm}}r@{\hspace{1.5mm}}r@{\hspace{1.5mm}}r
@{\hspace{1.5mm}}r@{\hspace{1.5mm}}r}
\hline
WR     & $\gamma_e$&   $Z$ &$\mu$ &  $\gamma_{{\rm S}^{2+}}$ &  
$\gamma_{{\rm S}^{3+}}$ & S/He &
$\gamma_{{\rm Ne}^+}$ &  $\gamma_{{\rm Ne}^{2+}}$ & Ne/He \\
       &           &       &      &   10$^{-5}$ &   10$^{-5}$ 
             & 10$^{-5}$ &  10$^{-4}$    & 10$^{-4}$     & 10$^{-4}$ \\
\hline
11 & 1.137 & 1.189 & 5.14 &1.9 & 3.1& 5.9  &5.4 &21.6& 31.9  \\ 
   &       &       &      &2.5 & 5.1& 9.0  &6.5 &27.8 &40.5  \\
90 & 1.219 & 1.287 & 5.84 &-- & --  & --  &$\le$4.0 & 22.6& $\le$34.0 \\ 
   &       &       &      &-- & --  & --  &$\le$4.8& 29.1& $\le$43.4\\
135& 1.138 & 1.189 & 5.21 &-- & --  & --  &$\le$5.9& 32.0& $\le$44.0 \\
   &       &       &      &-- & --  & --  &$\le$7.0& 41.1&$\le$55.8 \\
146& 1.083 & 1.117 & 4.70 &-- & --  & --  &$\le$2.8& 22.8&$\le$28.2 \\ 
   &       &       &      &-- & --  & --  &$\le$3.3& 29.5&$\le$36.1  \\ 
\hline
\end{tabular}
\end{center}
\end{table}

\begin{figure}
\vspace{12.75cm}
\includegraphics{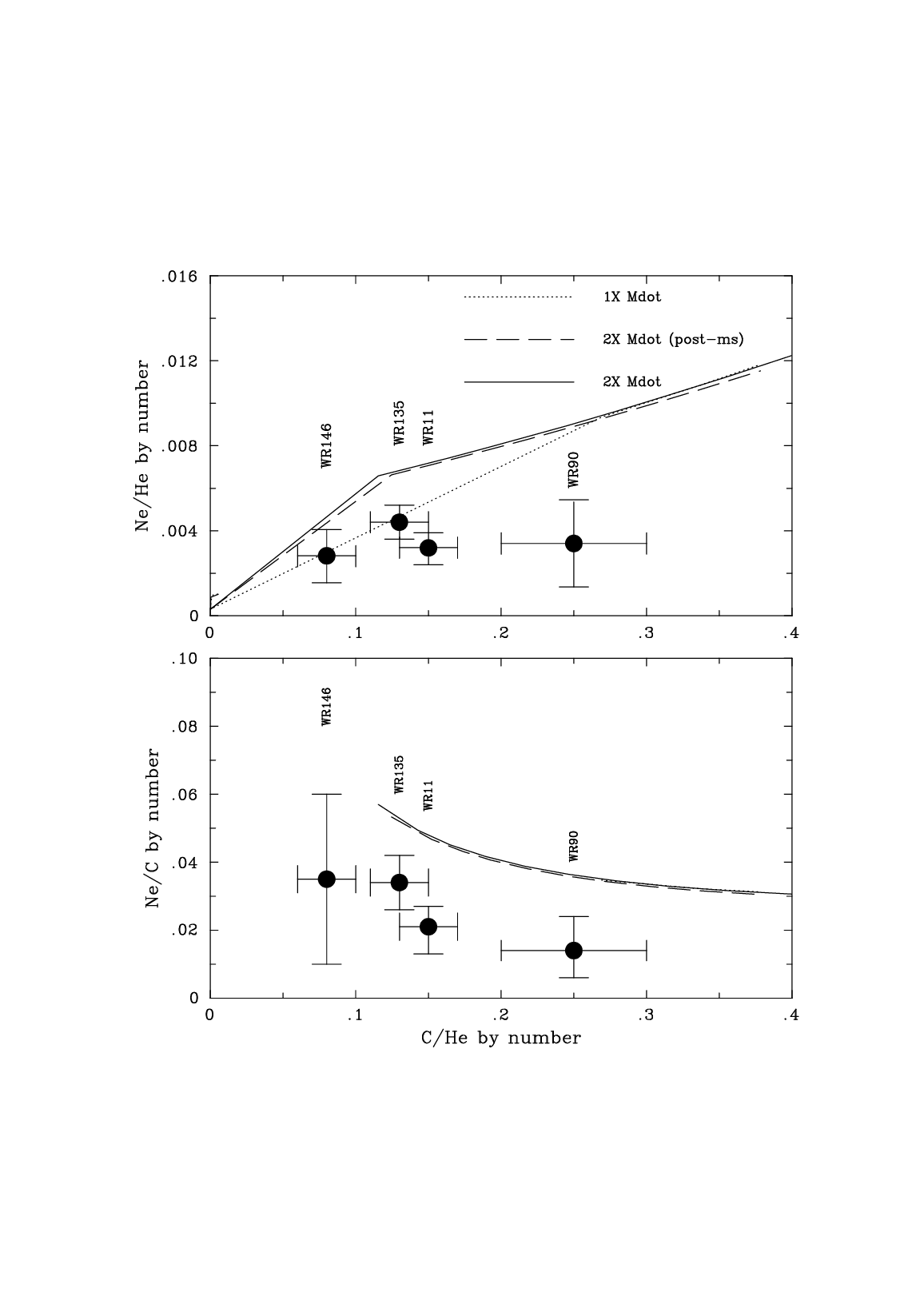}
\caption{Comparison between Ne/C (bottom) or Ne/He (top) ratios versus 
measured C/He, also confronted to theoretical
expectations from the solar metallicity evolutionary tracks 
of Schaller et al. (1992) and Meynet et al. (1994) for 60$M_{\odot}$. 
Three tracks are shown: (i) normal mass-loss rates during 
the entire evolution of the star (dotted); (ii) 
enhanced (2$\times$) mass-loss during the post main-sequence phase 
(dashed); (iii) enhanced (2$\times$) mass-loss during the entire stellar 
evolution (solid). Error bars account for uncertainties in distance and filling
factor (see text)}
\label{fig5}
\end{figure}

\subsection{Comparison of abundances with theoretical predictions}

Regarding theoretical expectations, the level of neon enrichment
is expected to be strongly correlated with the carbon content (Schaller
et al. 1992; Meynet et al. 1994). In Fig.\ref{fig5} 
we compare observed Ne/He and Ne/C versus C/He ratios with theoretical 
expectations from the Schaller et al. (1992) and
Meynet et al. (1994) evolutionary models at solar metallicity
for an initial mass of 60$M_{\odot}$. We find that observed neon abundances
are a factor of two below expectations. Specifically, we derive 
Ne/C=0.02$\pm$0.01 in all cases,
while Ne/C$\ge$0.03 is predicted during this phase of the WC evolution.
Error bars shown in the figure account for uncertainties in distance 
and (uniform) filling factors, but do not allow for the possibility of 
a varying volume filling factor between the neon and 
UV/optical line forming regions, which  
could be responsible for the discrepant cases. 

\section{Summary}\label{sect8}

We have performed quantitative analyses of a small sample of WC5--8 stars, 
using models that account for line blanketing and clumping. 
Comparisons
between synthetic spectra and de-reddened UV to mid-IR observations are
excellent, with few modelling deficiencies identified. Stellar parameters
support previous determinations (e.g. Koesterke \& Hamann 1995), except 
that the incorporation of blanketing yields higher stellar luminosities,
while clumping indicates lower wind performance numbers, supporting the
conclusions of Hillier \& Miller (1999) for WR111 (WC5). Future studies will 
derive properties of WC-type stars, at both earlier (WO)
 and later (WC9) spectral type, and investigate whether predicted ionizing
properties are consistent with nebular observations (e.g. Crowther et al. 
1999).

ISO/SWS spectroscopy reveals the presence of neon fine structure 
transitions, allowing abundance determinations. Using a revised
formulation of the BRA88 technique to account for wind clumping,
we derive neon abundances of Ne/He=3--4$\times$10$^{-3}$ by number,
seven times higher than the cosmic value adjusted for the H-depleted
WC environment, supported by Ne/S=50 for WR11
from sulphur fine structure lines. The Ne enrichment is a factor 
of $\sim$2 times lower than predictions of current theoretical models. 
However, differences in volume filling factors between the (high density) UV/
optical line formation regions and 
(low density) mid-IR fine-structure forming regions represent the 
greatest source of uncertainty in current Ne/He abundance determinations.
Nevertheless, Ne/S provides an independent confirmation of the neon enrichment
since it is {\it independent} of outer wind filling factors. 
Future large ground and space-based telescopes that are optimised for 
the IR, will allow neon and sulphur line flux measurements and abundance 
determinations for more distant Wolf-Rayet stars. Of particularly interest 
are abundances in carbon and oxygen-rich WO stars.

\section*{Acknowledgments}
This work is based on observations with ISO, an ESA 
project with instruments funded by ESA Member States (especially the PI 
countries: France, Germany, the Netherlands and the United Kingdom) with 
the participation of ISAS and NASA. 
Theoretical predictions presented here were possible only as a result of
the Opacity Project, led by Prof Michael Seaton. Thanks to Dr 
Tim Harries for observing WR90 on our behalf at the AAT. 
LD would like to acknowledge financial support from the UCL 
Perren Fund. PAC acknowledges financial support from a Royal Society 
University Research Fellowship. DJH gratefully acknowledges support by NASA 
through grant number NAG5--8211. The Anglo-Australian Telescope, Isaac Newton 
Telescope and U.K. Infrared Telescope are operated by the Anglo-Australian 
Observatory, Isaac Newton Group and Joint Astronomy Centre, respectively, 
on behalf of the Particle Physics and Astronomy Research Council.

\label{lastpage}


\begin{thebibliography}{}
\bibitem{} Abbott D.C., Bieging J.H., Churchwell E.  1981, ApJ 250, 645
\bibitem{} Abbott D.C., Bieging J.H., Churchwell E., Torres A.V, 1986, ApJ, 303, 239
\bibitem{} Barlow M.J., Roche P.F,  Aitken D.A., 1988, MNRAS 232, 821 (BRA88)
\bibitem{} Becker S.R., Butler K., 1992, A\&A 265, 647
\bibitem{} Becker S.R., Butler K., 1995a, A\&A 294, 215
\bibitem{} Becker S.R., Butler K., 1995b, A\&A 301, 187
\bibitem{} Butler K., Zeippen C.J., 1994, A\&AS 108, 1
\bibitem{} Cardelli J.A., Clayton G.C., Mathis J.S., 1989, ApJ 345, 245
\bibitem{} Chapman J., Leitherer C., Koribalski B., Bouter R., Storey M., 1999, 
ApJ 518, 890
\bibitem{} Conti P.S., Aschuler W.R., 1971, ApJ 170, 325
\bibitem{} Crowther P.A., Bohannan B., 1997, A\&A 317, 532
\bibitem{} Crowther P.A., De~Marco O., Barlow M.J., 1998, MNRAS,  296, 367
\bibitem{} Crowther P.A., Pasquali A., De~Marco O., Schmutz W., Hillier D.J.,
de Koter, A., 1999, A\&A, 350, 1007
\bibitem{} de Graauw Th., Haser L.N., Beintema D.A., et al. 1996, A\&A, 315, L49
\bibitem{} Daly P.N., Beard S.M., 1992, SUN 27 (Rutherford Appleton Laboratory)
\bibitem{} De Marco O., Schmutz W., Crowther P.A. et al. 2000, A\&A submitted
\bibitem{} Dougherty S.M., Williams P.M., van der Hucht K.A., Bode M.F., 
 Davis R.J., 1996,  MNRAS, 280, 963 
\bibitem{} Dougherty S.M., Williams P.M., Pollacco D.L., 2000,  MNRAS in press
\bibitem{} Eenens  P.R.J,, Williams P.M., Wade R., 1991, MNRAS, 252, 300
\bibitem{} Eenens  P.R.J,, Williams P.M., 1992, MNRAS, 255, 227
\bibitem{} Fernley J.A., Seaton M.J., Taylor K.T., 1987, J Phys B 20, 6457
\bibitem{} Galavis M.E., Mendoza C., Zeippen C.J., 1995, A\&AS 111, 347
\bibitem{} Hillier D.J., 1987, ApJS 68, 947
\bibitem{} Hillier D.J., 1990, A\&A 231, 111
\bibitem{} Hillier D.J., 1991, A\&A 247, 455
\bibitem{} Hillier D.J., 1996, in: Vreux J.M., Detal A., Fraipont-Caro D.,
Gosset E., Rauw G. (eds.), Wolf-Rayet stars in the Framework of Stellar 
Evolution, Proc 33rd Li\`ege Int. Ast. Coll., Universit\'e de Li\`ege, p.509
\bibitem{} Hillier D.J., Miller D.L. 1998, ApJ 496, 407
\bibitem{} Hillier D.J., Miller D.L. 1999, ApJ 519, 354
\bibitem{} Howarth I.D., 1983, MNRAS, 203, 301
\bibitem{} Howarth I.D., Phillips, A.P., 1986, MNRAS 222, 809
\bibitem{} Howarth I.D., Schmutz W., 1992, A\&A 261, 503
\bibitem{} Howarth I.D., Murray J., Mills D., Berry D.S., 1998
SUN 50.21, (Rutherford Appleton Laboratory)
\bibitem{} van der Hucht K.A., Olnon F.M., 1985, A\&A 149, 17
\bibitem{} van der Hucht K.A., Conti P.S., Lundstr\"{o}m I., Stenholm B., 1981,
Space Sci. Rev. 28, 227
\bibitem{} van der Hucht K.A., Hidayat B., Admiranto, A.G., Supelli K.R., Doom C.,
A\&A 1988, 199, 217
\bibitem{} van der Hucht, K.A., Morris P.W., Williams P.M., et al. 1996, A\&A
315, L193
\bibitem{} van der Hucht, K.A., Schrijver H., Stenholm, B., et al. 1997, New Astronomy, 2, 245
\bibitem{} de Jager C., Nieuwenhuijzen H., van der Hucht K.A., 1988, A\&AS 72, 259
\bibitem{} Kessler M., Steinz J.A., Anderegg M.E., et  al., 1996, 
A\&A, 315, L27 
\bibitem{} Koesterke L.,  Hamann W.-R, 1995, A\&A, 299, 503
\bibitem{} Kurucz R.L., 1991, in: Philip A.G.D., Upgren A.R., Janes K.A. (eds.),
Precision Photometry: Astrophysics of the Galaxy, L. Davis Press, Schenectady, p.27
\bibitem{} Langer N., 1989, A\&A 220, 135
\bibitem{} Leitherer C. Champan, J.M., Koribalski B., 1997 ApJ 481, 898
\bibitem{LM} Lepine S., Moffat A.F.J., 1999, ApJ 514, 909
\bibitem{} Lundstr\"{o}m I.,  Stenholm B., 1984, A\&AS 58, 163
\bibitem{} Luo D., Pradhan A.K., Saraph H.E., Storey P.J., Yu Yan., 1989, J Phys B., 22, 389
\bibitem{} Luo D., Pradhan A.K., 1989, J Phys B., 22, 3377
%
\bibitem{} Massey P., 1984, ApJ 281, 789
\bibitem{} Meynet G., Maeder A., Schaller G., Schaerer D., Charbonnel C., 1994, A\&AS 103, 97
\bibitem{} Moffat A.F.J., Drissen L., Lamontagne R., Robert C., 1988, ApJ
334, 1038
\bibitem{} Moffat A.F.J. 1999, in Wolf-Rayet Phenomena in Massive Stars and 
 Starburst Galaxies, eds. K.A. van der Hucht, G. Koenigsberger \& P.R.J. Eenens,
 Proc. IAU Symp. No. 193 (San Francisco: ASP), 278
\bibitem{} Morris P.W., Brownsberger K.R., Conti P.S., Massey P., Vacca W.D., 1993, ApJ 412, 324
\bibitem{} Morris P.W., van der Hucht, K.A., Willis A.J., Williams P.M., 1998, Ap\&SS 255, 157
\bibitem{} Morris P.W., van der Hucht, K.A., Crowther P.A., Dessart L., Willis
A.J., Williams P.M., 2000, A\&A 353, 624
\bibitem{} Morris P.W., van der Hucht, K.A., et al. 2000, A\&A in prep.
\bibitem{} Naqvi A.M., 1951, Thesis, Harvard
\bibitem{} Niemela V., Shara M.M., Wallace D.J., Zurek D.R., Moffat A.F.J.,  1998 AJ 115, 2047
\bibitem{} Nussbaumer H., Storey P.J., 1983, A\&A 126, 75
\bibitem{} Nussbaumer H., Storey P.J., 1984, A\&AS 56, 293
\bibitem{} Peach G., Saraph H.E., Seaton M.J.,  1988 J Phys B. 21, 3669
\bibitem{} Prinja R.K., Barlow M.J., Howarth I.D., 1990, ApJ 361, 607
\bibitem{} St Louis N., 1990, PhD thesis, University of London
\bibitem{} St-Louis N., Willis A.J., Stevens I.R., 1993, ApJ 415, 298
\bibitem{} Saraph H.E., Tully, J.A. 1994, A\&AS 107, 29
\bibitem{} Saraph H.E., Storey P.J., 1999, A\&AS 134, 369
\bibitem{} Schaeidt S., Morris, P.W., Salama, A., et al. 1996, A\&A, 315, 
L60 
\bibitem{} Schaerer D., Maeder A., 1992, A\&A 263, 129
\bibitem{} Schaerer D., Schmutz W., Grenon M., 1997 ApJ 484, L153
\bibitem{} Schaller G., Schaerer D., Meynet G., Maeder, A., 1992, A\&AS 96, 269
\bibitem{} Schmutz W.  1997, A\&A 321, 268
\bibitem{} Schmutz W., Vacca W.D., 1991, A\&AS 89, 259
\bibitem{} Seaton M.J., 1987, J. Phys B., 20, 6363
\bibitem{} Seaton M.J., 1995, The Opacity Project Volume 1, Institute of
Physics Publishing, Bristol.
\bibitem{} Shortridge K., Meyerdierks H., Currie M., et al. 1999, SUN 86.17 (Rutherford Appleton Laboratory)
\bibitem{} Smith L.F., 1968, MNRAS 140, 409
\bibitem{} Smith L.F., Shara, M.M., Moffat, A.F.J., 1990, ApJ 358, 229
\bibitem{} Steenman H., Th\'{e} P.S., 1989, Ap\&SS 159, 189
\bibitem{} Steenman H., Th\'{e} P.S., 1991, Ap\&SS 184, 9
\bibitem{} Stevens I.R. Corcoran M.F., Willis A.J. et al. ,1996, MNRAS, 283, 589
\bibitem{} Torres-Dodgen A.V., Massey P., 1988, AJ 96, 1076
\bibitem{} Torres-Dodgen A.V., Carroll M., Tapia M., 1991, MNRAS 249, 1
\bibitem{} Tully J.A., Seaton M.J., Berrington K.A., 1990, J Phys B 23, 3811
\bibitem{} Wiese W.L., Smith M.W., Glennon B.M., 1966, Atomic Transition
Probabilities, Volume I Hydrogen Through Neon, NSRDS-NDS 4.
\bibitem{} Willis A.J., van der Hucht K.A., Garmany C.D., Conti P.S., 1986, A\&AS, 63, 417
\bibitem{} Willis A.J., Dessart L., Crowther P.A., et al. 1997,  MNRAS 290, 371 (Paper~I)
\bibitem{} Wright A.E., Barlow M.J., 1975, MNRAS, 170, 41 
J Phys B 22, 389
\bibitem{} Yu Yan, Taylor K.T., Seaton M.J., 1987,  Phys B. 20, 6399
\bibitem{} Yu Yan, Seaton M.J., 1987,  Phys B. 20, 6409

\end{thebibliography}
\end{document}